%% file: main.tex
\documentclass{egpubl}
\usepackage{eg2022}

\STAR                   
\usepackage[T1]{fontenc}
\usepackage{dfadobe}  

\usepackage{cite}  
\BibtexOrBiblatex
\electronicVersion
\PrintedOrElectronic
\ifpdf \usepackage[pdftex]{graphicx} \pdfcompresslevel=9
\else \usepackage[dvips]{graphicx} \fi

\usepackage{egweblnk}

\title[A Survey on Reinforcement Learning Methods in Character Animation]%
      {A Survey on Reinforcement Learning Methods\\ in Character Animation}

\author[A. Kwiatkowski et al.]
{\parbox{\textwidth}{\centering Ariel Kwiatkowski$^{1}$, Eduardo Alvarado$^{1}$, Vicky Kalogeiton$^{1}$, C. Karen Liu$^{2}$, Julien Pettré$^{3}$, Michiel van de Panne$^{4}$ Marie-Paule Cani$^{1}$
        }
        \\
{\parbox{\textwidth}{\centering 
$^1$ LIX, École Polytechnique/CNRS, Institut Polytechnique de Paris, Palaiseau, France
\\
$^2$ Stanford University, Stanford, CA, USA
\\
$^3$ Univ Rennes, Inria, CNRS, IRISA, Rennes, France
\\
$^4$ University of British Columbia, Vancouver, Canada
      }
}
}

\usepackage{amssymb}
\usepackage{amsmath}
\usepackage{multicol}
\usepackage[utf8]{inputenc}

\usepackage[T1]{fontenc}
\usepackage[english]{babel}
\usepackage[utf8]{inputenc}

\usepackage{tikz}

\tikzset{
    vertex/.style = {
        circle,
        fill            = black,
        outer sep = 2pt,
        inner sep = 1pt,
    }
}

\newcommand{\RR}{\mathbb R}

\newcommand{\State}{\mathcal S}
\newcommand{\Action}{\mathcal A}
\newcommand{\MDP}{\mathcal M}
\newcommand{\EE}{\mathop{\mathbb{E}}}

\newcommand{\ie}{i.e.\ }
\newcommand{\eg}{e.g.\ }

\DeclareMathOperator*{\argmax}{arg\,max}

\begin{document}


\maketitle
\begin{abstract}
   Reinforcement Learning is an area of Machine Learning focused on how agents can be trained to make sequential decisions, and achieve a particular goal within an arbitrary environment. While learning, they repeatedly take actions based on their observation of the environment, and receive appropriate rewards which define the objective. This experience is then used to progressively improve the policy controlling the agent's behavior, typically represented by a neural network. This trained module can then be reused for similar problems, which makes this approach promising for the animation of autonomous, yet reactive characters in simulators, video games or virtual reality environments. This paper surveys the modern Deep Reinforcement Learning methods and discusses their possible applications in Character Animation, from skeletal control of a single, physically-based character to navigation controllers for individual agents and virtual crowds. It also describes the practical side of training DRL systems, comparing the different frameworks available to build such agents.


\end{abstract}

\input{text/01-introduction}

\input{text/02-preliminaries}

\input{text/03-classification}

\input{text/04-single-algo}

\input{text/05-multi-algo}

\input{text/06-single-appli}

\input{text/07-multi-appli}

\input{text/08-interaction}

\input{text/09-frameworks}

\input{text/10-conclusions}

\bibliographystyle{eg-alpha} 
\bibliography{text/references}       



\end{document}

%% file: text/01-introduction.tex
\section{Introduction}

Computer Graphics (CG) and Virtual Reality (VR) applications, from movies to video games, make a wide use of virtual characters, i.e. digital representations of humans, animals or other living creatures. For a long time, animation pipeline standards have pursued realism and control over motion style through fully kinematic characters, often designed manually by artists specifically for each situation, resulting in high time and resource costs. However, the increasing complexity of many applications makes the development of more versatile authoring tools a priority. In particular, simulators, games and VR environments share the need for autonomous characters, able to act in the expected way, while being reactive to any changes in their environment due to the user's actions. In order to produce such systems, learning-based approaches need to be explored.

Modern Machine Learning is commonly divided into three categories: Supervised Learning (SL), Unsupervised Learning (UL), and Reinforcement Learning (RL). Supervised Learning refers to learning using data with labels, Unsupervised Learning, including Self-Supervised Learning makes use of raw data without labels, and Reinforcement Learning does not use data in the usual sense.
Instead, the learning stage in RL consists of an agent taking a sequence of actions in one or more environments, and trying to maximize a reward function dependent on the states it visits.
During this process, the agent progressively trains its own controller module, which in the case of Deep Reinforcement Learning (DRL) is represented by a deep neural network. Once learned, the network can be used in a new, and possibly evolving environment, to make the agent take actions in a successful way towards its goals.


RL stands out as a promising approach for character animation because it provides a versatile framework to learn motor skills without the need of labelled data. RL is particularly useful when the dynamic equations of the environment are unknown or non-differentiable, to which conventional gradient-based optimal control algorithms do not apply.

Compared to traditional methods in AI, the designer does not need to specify what the character should do in each case -- a time-consuming and non scalable method. In contrast, the agent will discover the appropriate actions during the learning stage, given the targeted task or goals expressed in the form of a reward function.

This survey reviews the most common modern \textbf{DRL} algorithms, and how they can be used to tackle the main challenges in \textbf{character animation}. We consider two main categories of tasks -- individual motion skills, and motion planning tasks. Individual scenarios typically involve skeletal motion control of a physically-based character, while motion planning often involves multiple characters interacting in a shared environment. In particular, we focus also on the problem of \textbf{crowd simulation}, which focuses on determining the trajectories of multiple agents in a shared environment, often abstracting away their internal structure.


We begin by describing the main, most recent challenges in the field of character animation (Sec.~\ref{ssec:problems}). Then, we present the key principles and notations in RL (Sec.~\ref{sec:definitions}), and continue with a general classification of the most common approaches (Sec.~\ref{sec:classification}). Subsequently, we divide the addressed RL solutions into two groups: single-agent (Sec.~\ref{sec:single-algo}) and multi-agent (Sec.~\ref{sec:multi-algo}) problems. Then, we describe how these methods are used to solve computer animation problems, for skeletal motion control (Sec.~\ref{sec:applications}) and navigation problems (Sec.~\ref{sec:appli-crowd}), as well as some works concerning interactions between virtual agents and humans (Sec.~\ref{sec:interaction}). Finally, we present a description of current, available frameworks to apply RL-based solutions (Sec.~\ref{sec:frameworks}), before concluding with a summary of the most relevant algorithms used for a particular problem.

Our work is largely complementary to a recent survey on deep learning for skeleton-based human animation~\cite{mourot2021survey},
which we also recommend to readers.  In particular, we provide a detailed review of current RL methods (both single agent and multiagent) 
and their mathematical foundations, a full review of RL methods for character navigation methods, and a complementary classification of physics-based character RL methods.

\subsection{Problems in Character Animation} \label{ssec:problems}

In the most general sense, the field of \textbf{Character Animation} concerns everything related to  animating virtual characters. In this work specifically, we focus on the aspects of \textbf{behavior} of said agents, on their skeletal motion control, as well as on their \textbf{interactions} with a possible human user. Topics related to modeling and animating the character's face, skin, muscles, hair and clothes, or rendering it are out of scope of this report.

When dealing with a single animated character (which may also encompass situations with several independent characters), there are two main levels that need to be considered:
\begin{itemize}
    \item Skeletal Animation
    \item Character Motion Planning
\end{itemize}

Skeletal Animation deals with internal motions of an agent -- how the individual limbs move, while the position in the global frame may be of secondary concern. Character Motion is the opposite -- it abstracts away the details of the character's shape, instead focusing on its displacements through the scene.

When considering Character Motion for multiple interacting characters, the problem turns into that of Crowd Simulation. Typically, in those problems, each agent has a destination it wants to reach, while avoiding collisions with the environment and with other agents. Van Toll and Pettré~\cite{toll_algorithms_2021} wrote an overview of the modern approaches from the last decade.

%% file: text/02-preliminaries.tex
\section{Definitions and Preliminaries} \label{sec:definitions}

In this section, we introduce the basic formal background of Reinforcement Learning. First, we describe and compare different ways of formalizing the RL task to specify \textbf{what} we want to solve. Then, we describe the fundamental theorems supporting modern RL methods to show \textbf{how} we can solve those tasks.

\subsection{Reinforcement Learning Formalisms}

\begin{figure}
    \centering
    \includegraphics[width=\linewidth]{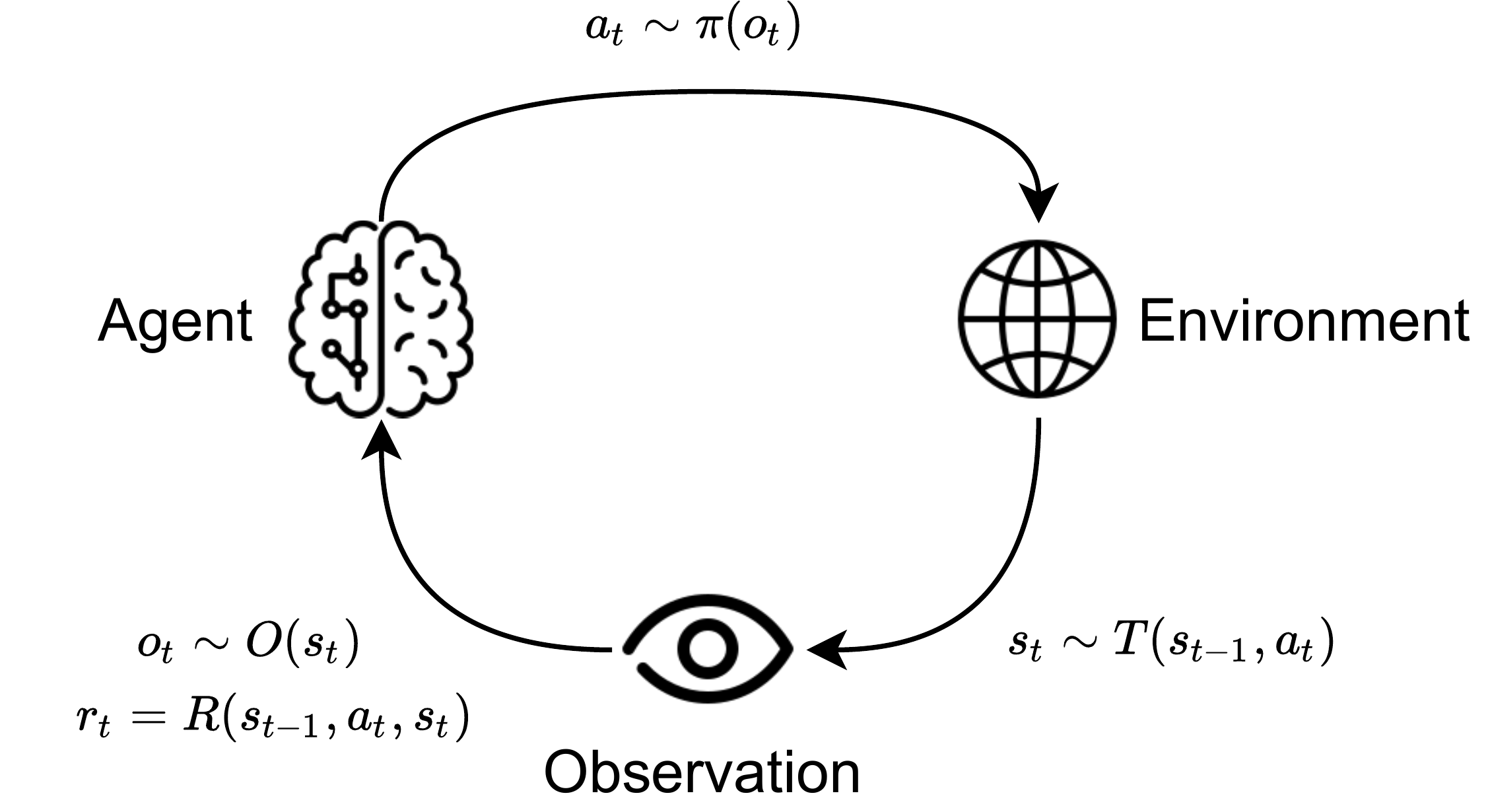}
    \caption{A visual depiction of the basic Reinforcement Learning loop corresponding to the POMDP formalism. The agent and the environment exchange information between each other. The agent perceives the environment state and executes an action. The environment then updates its state, and communicates it to the agent via an observation function, together with the reward for the last action.}
    \label{fig:rl-loop}
\end{figure}

While there exist several frameworks that are used to formalize the Reinforcement Learning problem, they are based on the Markov Decision Process~\cite{bellman_markovian_1957, sutton_introduction_1998, sutton_reinforcement_2018} (MDP), with variations adapting it to the specific task at hand. In this section, we describe the variants relevant to character animation, both for individual agents, as well as multiagent scenarios.

In essence, a Reinforcement Learning problem consists of two parts -- an \textbf{environment}, and an \textbf{agent} acting within that environment in order to achieve some goals. The agent observes the environment, receiving its \textbf{state} or \textbf{observation}, and based on that executes an \textbf{action}. The state of the environment then changes, and the agent receives a reward signal indicating how good that action was. The agent's objective is maximizing the total reward collected during an episode. An episode starts from an initial state, and lasts until the agent reaches a terminal state, or the environment terminates otherwise (\eg due to a time limit). A schematic representation of this loop is in Figure~\ref{fig:rl-loop}.

\subsubsection{Single Agent} \label{sec:single-mdp}

A general \textbf{Markov Decision Process} (\textbf{MDP}) is defined by a tuple $\MDP = (\State, \Action, T, R, \mu)$, optionally with a sixth component $\gamma$ (which can also appear in all other formalisms, and hence will be omitted from their descriptions), where:
\begin{itemize}
    \item $\State$ is a set of states of the environment.
    \item $\Action$ is a set of actions available to the agent.
    \item $T\colon \State \times \Action \to \Delta \State$ is the environment transition function, representing its dynamics.
    \item $R\colon \State \times \Action \times \State \to \RR$ is the reward function which is used to define the agent's task.
    \item $\mu \in \Delta \State$ is the initial state distribution.
    \item $\gamma \in [0, 1]$ is the discount factor.
\end{itemize}
Note that we use the notation $\Delta X$ to represent for the set of all probability distributions over the set $X$.

During an episode, an initial state $s_0 \in \State$ is sampled from $\mu$. The state is typically represented by a continuous vector in $\RR^n$, or in simple cases, a discrete value. After which the agent repeatedly selects an action $a_t$ from $\Action$, observes a new state $s_{t+1} \sim T(s_t, a_t)$ and receives a reward $r_t = R(s_t, a_t, s_{t+1})$. The actions, similarly to observations, are typically continuous vectors or discrete values, although more complex nested structures are also used. This can repeat infinitely, or until some termination condition, defined either by a terminal state in $\State$, or a time limit. The agent's objective is maximizing the \textbf{total discounted reward} $\sum_t \gamma^t r_t$, or simply non-discounted \textbf{total reward} $\sum_t r_t$ when $\gamma=1$.

The solution to an MDP is defined as an \textbf{optimal policy}, typically denoted as $\pi^*\colon \State \to \Delta \Action$. It is the policy that, when executed, leads to the highest expected total discounted reward. While a policy may be stochastic or deterministic, depending on the properties of the action distributions it outputs, note that the optimal policy is generally stochastic, \ie it returns a distribution over actions rather than a specific action.
For consistency, the notation we use in this work is that the action distribution of a policy $\pi$ in a given state $s$ is $\pi(s)$, whether that policy is stochastic or not. The action is then sampled from the policy $a \sim \pi(s)$. An alternative notation uses the notion of a conditional probability of the action given the current state $\pi(a|s)$, and is equivalent to ours.

A key property of MDPs is their full observability - agents have complete information of the current environment state. This is rarely the case in real applications, and thus a \textbf{Partially Observable Markov Decision Process}~\cite{kaelbling_planning_1998} (\textbf{POMDP}) is often used instead. 

A POMDP is defined by a tuple $\MDP = (\State, \Action, T, R, \Omega, O, \mu)$, where $\State, \Action, T, R, \mu$ are defined as in MDPs. $\Omega$ is a set of possible observations, and $O\colon S \to \Delta \Omega$ is the observation function mapping states to observations. This time, the agent does not perceive the real state $s_t$ of the environment, but rather the observation $o_t \sim O(s_t)$ which may not contain the full information, hence the partial observability.

\subsubsection{Multiagent} \label{sec:multi-mdp}


While the MDP and POMDP formalisms are sufficient for problems with a single agent, the generalization to multiple agents can be done in different ways depending on the extent of flexibility required for a given application. The most general case is a \textbf{Partially Observable Stochastic Game}~\cite{hansen_dynamic_2004} (\textbf{POSG}) which is defined as a tuple $\MDP = (\mathcal{I}, \State, \{ \Action_i \}, \{\Omega_i\}, \{O_i\}, T, \{R_i\}, \mu)$, where:

\begin{itemize}
    \item $\mathcal{I}$ is the finite set of agents, indexed $1, \dots, n$
    \item $\State$ is a set of states of the shared environment.
    \item $\Action^i$ is a set of actions available to agent $i$, and $\Action = \times_{i \in \mathcal{I}} \Action^i$ is the joint action set.
    \item $\Omega^i$ is the set of observations available to agent $i$.
    \item $O^i\colon \State \to \Omega_i$ is the observation function for agent $i$.
    \item $T\colon \State \times \Action \to \Delta \State$ is the environment transition function, representing its dynamics.
    \item $R^i\colon \State \times \Action \times \State \to \RR$ is the reward function of agent $i$, which defines the agent's task.
    \item $\mu \in \Delta \State$ is the initial state distribution.
\end{itemize}

Similarly to the single-agent scenario, the environment is initialized with a state $s_0$ sampled from $\mu$. Each agent then receives an observation $o_t^i = O_i(s_t)$ and based on that, chooses an action $a_t^i$. The environment changes according to the joint action of all agents $a_t = (a_t^1, a_t^2, \dots, a_t^n)$ generating the new state $s_{t+1}$, and each of them then receives their rewards $r_t^i = R_i(s_t, a_t, s_{t+1})$ and observations $o_{t+1}^i$. This repeats until the episode ends. Since each agent receives its own reward, the objective of an agent $i$ is maximizing its total reward $\sum_t r_t^i$.


A special case of POSG is a \textbf{Decentralized Markov Decision Process}~\cite{bernstein_complexity_2000} (\textbf{DecPOMDP}) in which all agents work together to optimize a shared reward function $\forall_i R_i = R$. This formalism is suitable for fully cooperative tasks. It is worth noting that any POSG can be converted into a POMDP by setting the reward to be equal to the sum of of individual rewards, but it will not make sense in all POSGs (consider for example any zero-sum game).


An alternative, but equivalent to POSG formulation, is the \textbf{Agent Environment Cycle Game}~\cite{terry_agent_2021} (\textbf{AEC}) formalism. As opposed to the previous options, it is more adapted to dealing with environments in which agents do not act simultaneously. Formally, an AEC is defined by a tuple $\MDP = (\mathcal{I}, \State, \{ \Action_i \}, \{\Omega_i\}, \{O_i\}, P, \{T_i\}, \{R_i\}, \nu, \mu)$, where $T_i\colon \State \times \Action_i \to \State$ is a deterministic agent transition function, $P\colon \State \to \Delta \State$ is the environment transition function, $\nu\colon \State \times \mathcal{I} \times \Action \to \Delta \mathcal{I}$ is the \textit{next agent} function which determines which agent will be taking the action next. The other symbols are defined as before, with the exception of $\mathcal{I}$ which now also includes environment itself considered as a separate agent, represented by the symbol $0$. Furthermore, $\Action$ is now an union of all individual action spaces. All agents, including the environment, take turns taking their actions and modifying the shared state, which enables a greater flexibility compared to the POSG formalism. AEC environments are primarily used in the Petting Zoo framework~\cite{terry_pettingzoo_2021} (see Section \ref{sec:frameworks}).


In some cases, a more game theory-based approach is useful. The \textbf{Extensive Form Game}~\cite{lanctot_openspiel_2020} (\textbf{EFG}) formalism is notably used in the OpenSpiel framework. It contains implementations of many board games, which is the context that it excels in. However, it is not very applicable to character animation, and thus we refer the reader to the associated paper for further details on this formalism.

\textbf{Note}: Many details of the described formalisms vary between sources in the ordering of their elements, the size of the tuple, and the signatures of the functions. This does not change the underlying behavior, and we will therefore omit discussing the different descriptions of the same formalism.

\subsubsection{Environment design}

A crucial element when applying RL to new problems is designing an appropriate environment. This often involves building a simulation that implements the common API of Gym (see Section~\ref{sec:frameworks-envs}), since a purely mathematical formulation would quickly become very convoluted in a complex scenario. Note that we omit the transition function from this description, as this is typically part of the underlying simulation, and can therefore be implemented in any way.

The first consideration is the observation space. This is commonly represented as a fixed-size vector space $\RR^n$, which can be directly used with regular feed-forward neural networks. More complex nested structures as well as images are also possible, but they require an adaptation in the structure of the policy being learned. 

Second comes the action space. Depending on the environment, a common choice is either a vector space $\RR^n$, or simply a finite set of actions $|\Action| = n < \infty$. While from the point of view of the implementation it is important that the action space remains constant between different states, one can employ invalid action masking to restrict the available actions to a specific subset. Similarly to observations, it is also possible to use nested structures as long as the policy is adapted correspondingly.

Finally, the reward function defines the actual task and guides the agent's behavior. This is often the most critical component to develop, as a misspecified reward function can lead to unexpected and undesirable behaviors. The simplest reward function can be obtained by choosing a goal state, and giving the agent a reward of $1$ if it reaches that state, or $0$ otherwise. However, this sparse reward tends to make it very difficult for the agent to learn, as it needs to reach it with random exploration to receive any training signal. A common method is then using reward shaping~\cite{ng_policy_1999} by adding a smaller, dense reward that guides the agent towards the goal. In other cases, there might be a natural dense reward that can be used instead of the sparse one, such as the distance from the goal in environments with relatively simple dynamics.

\subsubsection{Summary}

\input{tables/formalisms}

We presented the most commonly used formalisms underlying the RL problem, which serve as a basis for finding ways to solve these tasks. The similarities and differences between them are in Table~\ref{tab:formalisms}. 
Typically, either MDP or POMDP can be used with a single agent. POMDP offers stronger theoretical justification if the agent does not observe the full environment state, but this high rigor is not always necessary. Instead, MDP is often used due to its simplicity. With multiple agents, POSG is a versatile choice that can work with any scenario.
If one needs to put an emphasis on some aspect of the environment, other options are also available. It is worth noting that those formalisms have dynamic programming solutions associated with them for cases with discrete action and state spaces. This however is impractical in complex scenarios that emerge in character animation, requiring more sophisticated algorithms.

\subsection{Fundamentals of RL Algorithms}

In this section we describe the mathematical theorems underlying the most important RL algorithms used today. Specifically, we show how the \textbf{Policy Gradient Theorem} enables directly optimizing a behavior policy function, and the \textbf{Bellman Equation} enables learning the expected utilities of actions that the agent can take in a certain state. These will serve as a basis for many modern algorithms, which often combine the two aspects. We use the notation of MDPs described in Section \ref{sec:single-mdp} because they provide sufficient generality. Under partial observability, states are replaced with observations, and multiagent extensions of relevant algorithms are discussed in Section \ref{sec:multi-algo}.

In both cases, modern algorithms use Neural Networks as approximators for the relevant functions. Because a detailed explanation of training neural networks is out of the scope of this work, we refer the readers to \eg the Deep Learning Book~\cite{goodfellow_deep_2016} for more information on that topic.

\subsubsection{Policy Gradients} \label{sec:policy-gradient}

The Policy Gradient Theorem is a basis for all \textbf{Policy Gradient} (\textbf{PG}) algorithms, starting with the seminal REINFORCE algorithm~\cite{sutton_policy_1999}. In the context of deep reinforcement learning, the policy $\pi\colon \State \to \Delta \Action$ is represented  as a neural network, and its free parameters, e.g., the weights, are optimized using gradient ascent on the total expected reward. In order to do that, we need to find the gradient with respect to the network's weights using a batch of collected experiences. Here we show a proof of the theorem based on that published in OpenAI Spinning Up~\cite{achiam_spinning_2018}, although other approaches for proving the same result exist~\cite{williams_simple_1992, jones_clearer_2020}.

Consider a trajectory in the environment, defined as a sequence of consecutive states and actions taken by the agent, and rewards $\tau = (s_0, a_0, r_0, s_1, a_1, r_1 \dots)$. Given the parametrized policy $\pi_\theta$, we know that the probability of a trajectory is

\begin{align}
    P(\tau) &= \mu(s_0) \prod_t P(s_{t+1}|s_t, a_t) \pi_\theta(a_t|s_t) \\
    \log P(\tau) &= \log \mu(s_0) + \sum_t \left( \log P(s_{t+1}|s_t, a_t) + \log \pi_\theta(a_t|s_t) \right)
\end{align}
and the total reward obtained in the trajectory is $R(\tau) = \sum_t r_t$


Consider now the expectation across all trajectories $\tau$. With the optimization target defined as $J(\theta) = \EE_{\tau\sim \pi_\theta} R(\tau)$, using a few calculus transformations, we can express the policy gradient as:

\begin{align}
    \nabla_\theta J(\theta) &= \nabla_\theta \EE_{\tau\sim \pi_\theta} R(\tau) \\
    &= \nabla_\theta \int_\tau P(\tau|\theta) R(\tau) \\
    &= \int_\tau \nabla_\theta P(\tau|\theta) R(\tau) \\
    &= \int_\tau P(\tau|\theta) \nabla_\theta \log P(\tau|\theta) R(\tau) \\
    &= \EE \left[ \nabla_\theta \log P(\tau|\theta) R(\tau) \right] \\
    &= \EE_{\tau\sim\pi_\theta} \left[ \sum_t \log \pi_\theta(a_t|s_t) R(\tau) \right]
\end{align} 

With this, given a batch of trajectories $\mathcal{D}$ collected using the policy we are optimizing, we can finally compute the gradient estimate:

\begin{equation}
    \hat{g} = \frac{1}{|\mathcal{D}|} \sum_{\tau \in \mathcal{D}} \sum_t \nabla_\theta \log \pi_\theta(a_t|s_t) R(\tau)
\end{equation}

Note that this is merely the base form of the theorem, and various modifications are possible, most notably in the form of \textbf{importance sampling}~\cite{rubinstein_simulation_1981}, or adding a baseline to the reward $R(\tau)$. Some of these are discussed in the context of specific algorithms that use them in Section~\ref{sec:single-algo}.

\subsubsection{Bellman Equation} 


The Bellman Equation~\cite{bellman_dynamic_2003} is a basis for all value-based algorithms. Unlike the Policy Gradient method, here we do not learn a policy directly. Instead, we try to approximate a state value function $V(s)$ or a state-action value function $Q(s, a)$. The former estimates the expected reward that the agent will collect in the future, given that it is present in a given state $s$. The latter estimates the same quantity, but given that the agent will take the specific action $a$ in the state $s$. Then, we use these functions to generate a policy by choosing the best action in a given state. With a state value function $V$, this requires access to the environment transition function, which is not necessary with a state-action value, where the policy is simply given by $a = \argmax_{a'} Q(s, a')$. 

The value function $Q^\pi$ (or analogously $V^\pi$) associated with a policy $\pi$ represents the expected total reward if the agents is in a given state $s$, takes a certain action $a$ ($a\sim\pi(s)$ for the state value function), and then proceeds by following the policy $\pi$ for the rest of the episode.

\begin{align}
    V^\pi(s) &= \EE\limits_{a_t\sim\pi} \left[ \sum_t \gamma^t r_t |  s_0 = s \right] \label{eq:v-definition}\\
    Q^\pi(s, a) &= \EE\limits_{a_t\sim\pi} \left[ \sum_t \gamma^t r_t | s_0=s, a_0=a\right] \label{eq:q-definition}
\end{align}

The $Q$ (or $V$) values of different state-action pairs (states) are obviously not independent -- they are in fact related via the transition function, which determines what state comes after them. This is formalized by the Bellman Equation, which defines the consistency criterion of a $Q$ (or $V$) function (Equations~\ref{eq:bellmanVpi}, \ref{eq:bellmanQpi}), and the optimal function $Q^{*}$ (or $V^*$) (Equations~\ref{eq:bellmanVopt}, \ref{eq:bellmanQopt}):

\begin{align}
    V^{\pi}(s) &= \EE_{\substack{a\sim\pi \\ s'\sim T}} \left[ R(s, a) + \gamma V^\pi(s') \right] \label{eq:bellmanVpi} \\
    V^{*}(s) &= \max_a \EE_{s'\sim T} \left[ R(s,a) + \gamma V^{*}(s') \right] \label{eq:bellmanVopt} \\
    Q^{\pi}(s, a) &= \EE_{s'\sim T} \left[ R(s, a) + \gamma \EE_{a'\sim\pi} Q^\pi(s', a') \right] \label{eq:bellmanQpi} \\
    Q^{*}(s, a) &= \EE_{s'\sim T} \left[ R(s,a) + \gamma \max_{a'} Q^{*}(s', a') \right] \label{eq:bellmanQopt}
\end{align}

The intuition behind these equations is that the value of a state is equal to the instant reward obtained at that state, and the discounted expected value of the following state -- which also includes the value of the state after that (due to the recursive nature of the equation), and so on until a terminal state. The value of a terminal state is typically considered to be $0$, however a different convention may be used in certain cases, \eg if the episode timed out. It also induces a dynamic programming solution of MDPs through the Value Iteration algorithm~\cite{sutton_reinforcement_2018}. It is however inapplicable or impractical for many modern problems with complex state and action spaces, and instead, the Bellman Equation is used as the source of a differentiable loss function for value-based algorithms, as we describe in detail in Section~\ref{sec:single-algo}.

It is worth noting that by using a $Q$ function estimator $\hat{Q}^\pi$, we can obtain an alternative formulation of the Policy Gradient Theorem. Indeed, as shown by Sutton et al.~\cite{sutton_reinforcement_2018}, we get the following expression for the policy gradient:

\begin{align}
    \hat{g} = \sum_s d^\pi(s) \sum_a \nabla \pi(a|s) \hat{Q}^\pi(s, a)
\end{align}

where $d^\pi(s)$ is the marginal state distribution under the policy $\pi$. This formulation does not use individual transitions, but instead relies on statistics of the policy's performance, and can thus be used as an alternative algorithm to estimate the policy gradient.

\subsection{Reward Hypothesis, Discounting, Advantage} \label{sec:rew-dis-adv}

It is worth taking a closer look at the assumption underlying all Reinforcement Learning research, sometimes called the \textbf{Reward Hypothesis}. It is formulated by Richard Sutton as ``That all of what we mean by goals and purposes can be well thought of as maximization of the expected value of the cumulative sum of a received scalar signal (reward)" \cite{sutton_reinforcement_2018}. This is reflected in the described formalisms and equations by the inclusion of a reward function $R$, with the goal of agents being maximization of the total reward obtained over their lifetime. Some argue that just the reward signal is sufficient to represent any goals that intelligent agents might have \cite{silver_reward_2021}, while others point out that certain objectives cannot be represented with a single scalar reward \cite{alexander_archimedean_2020}. That being said, as we focus specifically on Reinforcement Learning in this work, we do not consider alternative formulations -- but it is possible that they will become more relevant in the coming years as the field continues to develop.

An important element related to the reward function is the \textbf{discount factor} mentioned in the description of an MDP in Section~\ref{sec:single-mdp}. It can be considered either as a property of the environment, or the learning agent, and while the two views are mostly equivalent from the optimization point of view, they have potential implications relating to the Value Alignment problem~\cite{yampolskiy_value_2018}. If we consider the discount factor to be a property of the MDP, this is the real reward we want the agent to optimize, whereas otherwise, we really want to optimize the total reward, and discounting the rewards helps improve the training in some way, \eg as a form of regularization~\cite{amit_discount_2020}. It can also impact the range of methods that we can use -- when considered as a part of the learning agent, any arbitrary method of reward discounting can be used, including non-exponential methods such as hyperbolic~\cite{fedus_hyperbolic_2019} or truncated~\cite{lattimore_time_2011} discounting.

One issue with using the raw rewards/utility for training is that it is an absolute metric, with no a priori point of reference. If the agent only perceives a single timestep where a certain action $a_0$ leads to a reward of $-1$, this action's probability will be decreased as the value is negative. However, it could still be the optimal action if the counterfactual rewards due to taking other actions are even lower. Asymptotically, this is all balanced out due to the fact that decreasing the probabilities of other actions will necessarily increase the probability of $a_0$. To decrease the variance of gradient estimation, some algorithms use the notion of \textbf{Advantage} instead. This often results in more stable and efficient training. Intuitively, advantage measures how much a certain action is better (or worse) than expected. Given both the Q and V function approximations, we define the advantage as:

\begin{equation}
    A(s, a) = Q(s, a) - V(s)
\end{equation}

In practice, algorithms that use advantage often compute $Q(s, a)$ from collected experience, \ie, look at the trajectory and compute the total reward, while $V(s)$ is approximated with a separate neural network. Examples of this are included in Section~\ref{sec:single-algo}.



%% file: tables/formalisms.tex
\begin{table*}[ht] \centering
\caption{A comparison of different formalisms used to define an RL problem. 
Legend: $\times$ -- the property cannot be modelled in this formalism, $\thicksim$ -- the property can be modelled in this formalism, but is not the intended use or requires extra effort, $\checkmark$ -- the property can be modelled in this formalism, $\bigstar$ -- this formalism is particularly suitable for this property. Multiagent Cooperative and Competitive refers to the rewards being either shared or zero-sum, respectively. Multiagent Mixed is neither fully cooperative nor competitive. Simultaneous and Turn-based refers to whether all agents take their actions at the same time, or only one agent does.} \label{tab:formalisms}
\begin{tabular}{ccccccc}
Property                & MDP        & POMDP      & POSG         & DecPOMDP     & AEC          & EFG          \\ \hline
Single Agent            & $\bigstar$ & $\bigstar$ & $\checkmark$ & $\checkmark$ & $\checkmark$ & $\checkmark$ \\
Multiagent Cooperative & $\times$   & $\times$     & $\checkmark$ & $\bigstar$   & $\checkmark$ & $\checkmark$ \\
Multiagent Competitive  & $\times$   & $\times$   & $\checkmark$ & $\times$     & $\checkmark$ & $\checkmark$ \\
Multiagent Mixed        & $\times$   & $\times$   & $\bigstar$   & $\times$     & $\bigstar$   & $\bigstar$   \\
Multiagent Simultaneous & $\times$   & $\times$   & $\bigstar$   & $\bigstar$   & $\thicksim$  & $\thicksim$  \\
Multiagent Turn-based   & $\times$   & $\times$   & $\thicksim$  & $\thicksim$  & $\bigstar$   & $\bigstar$   \\
Partial Observability  & $\times$   & $\bigstar$   & $\checkmark$ & $\checkmark$ & $\checkmark$ & $\checkmark$ \\
Full Observability     & $\bigstar$ & $\checkmark$ & $\checkmark$ & $\checkmark$ & $\checkmark$ & $\checkmark$
\end{tabular}
\end{table*}

%% file: text/03-classification.tex
\section{Classification of RL Algorithms} \label{sec:classification}

\begin{figure*}
    \centering
    \includegraphics[width=\linewidth]{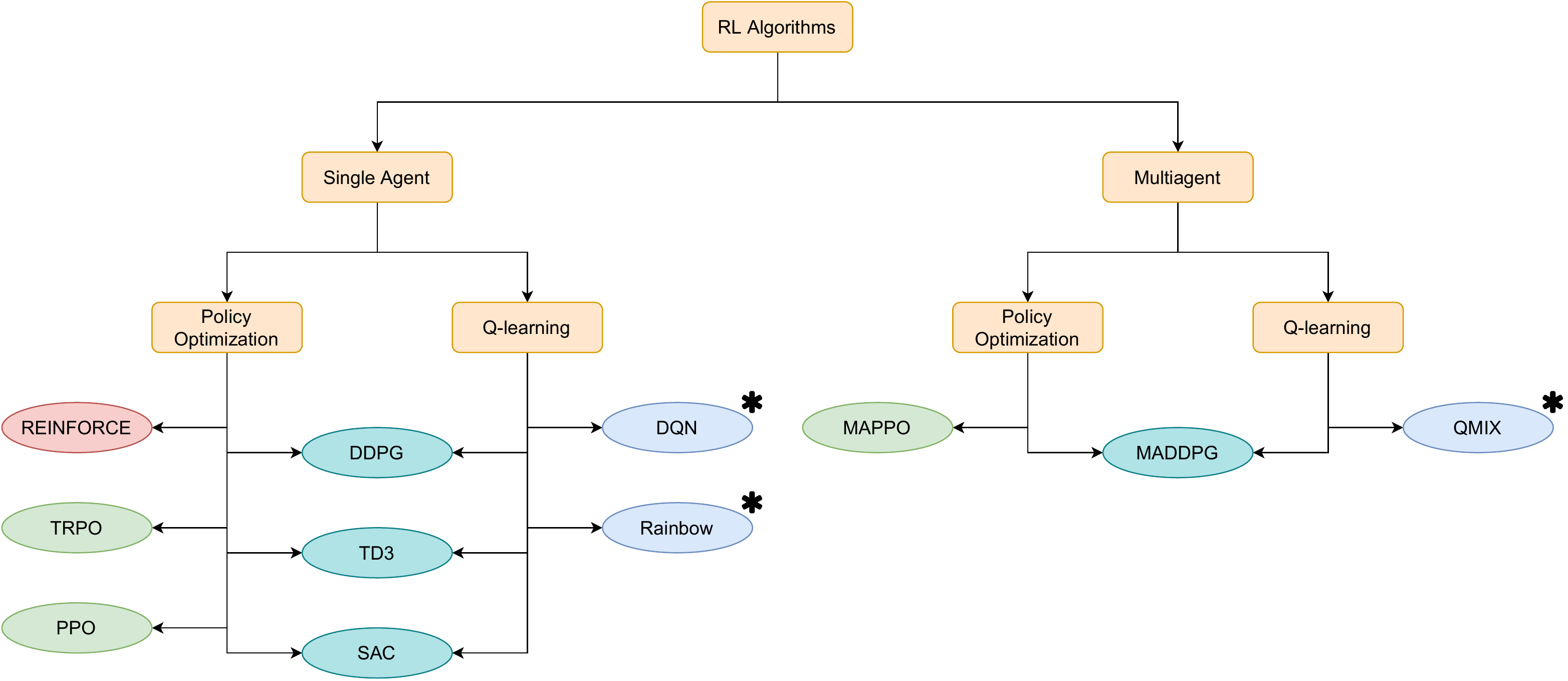}
    \caption{A diagram showing a taxonomy of the Reinforcement Learning algorithms described in this work. We focus on two divisions: single agent or multiagent, and policy-based or value-based. The colors of nodes correspond to whether the algorithm is on-policy (red), off-policy (blue), or in between (green). Algorithms marked with an asterisk (\pmb{\textasteriskcentered}) can only be used with discrete action spaces.}
    \label{fig:classification}
\end{figure*}

In this section we describe the main categories of modern RL algorithms. While the division is not clear-cut and many algorithms at least draw on ideas from other types, we nevertheless consider this classification to be useful for building an intuition of the RL algorithm landscape. A diagram classifying the algorithms described in this work is in Figure~\ref{fig:classification}. The details of these algorithms are provided in Sections~\ref{sec:single-algo} and \ref{sec:multi-algo}.

\subsection{Policy-based or Value-based}

The first axis of division is whether the algorithm is \textbf{policy-based (PB)} or \textbf{value-based (VB)}. Although the state-of-the-art algorithm often use both components via Actor-Critic architectures, oftentimes they still have one part that is dominant in the overall picture. The difference between PB and VB algorithms is in what the model is actually trained to predict. In pure PB algorithms such as REINFORCE~\cite{williams_simple_1992, sutton_policy_1999}, the neural network is trained to directly output the \textbf{action} that will maximize the expected future reward. On the other hand, pure VB algorithms like Deep Q Learn
ing (DQN)~\cite{mnih_human-level_2015} train the network to instead output the \textbf{value} of each action in a given state, that is the expected future reward. This works in environments with a discrete action space, because a policy can then be generated by taking the action with the maximum expected value. 

\subsection{Actor-Critic}

Very commonly, RL algorithms use the so-called Actor-Critic architecture, which involves training two networks. One, the \textbf{Actor}, also called the \textbf{policy}, is responsible for predicting the action that the agent should take, as in PB algorithms. The other, the \textbf{Critic}, is responsible for predicting the \textbf{value} of an action in a given state, as in VB algorithms. The outputs of the two networks, while not always in agreement with each other, can be used to improve the training process in ways that depend on the exact algorithm -- for example, by using the value prediction as a baseline for advantage estimation as in PPO~\cite{schulman_proximal_2017}, or by training the Actor to find the action with the highest value predicted by the Critic in order to use value-based methods in continuous action spaces as in DDPG~\cite{lillicrap_continuous_2015}.

\subsection{On-policy or Off-policy}


Another point of difference between RL algorithms is the data used for the optimization process, which does not necessarily have to be obtained with the same policy that is being learned. We normally refer to the \textbf{target policy} as the policy that is being optimized and will be used for evaluation, and the \textbf{behaviour policy} as the policy used by the agent to select actions and explore the environment. In \textbf{on-policy} algorithms like REINFORCE, the neural network can only be trained using data collected with the policy that is being optimized, meaning that the behavior policy matches the target policy. This implies that after performing a single gradient update, the data (in theory) has to be discarded. On the other hand, in \textbf{off-policy} algorithms like DQN, any data (trajectories) can be used, regardless of how it was generated (target and behaviour policies can be different). Some algorithms like PPO toe the line between being on-policy and off-policy, by allowing a relatively small number of gradient updates before the data has to be discarded by using tricks like importance sampling and clipping the loss function. Nevertheless, these algorithms are typically considered to be on-policy, as they cannot use data collected by an arbitrary behavior policy.

Typically, on-policy algorithms use a \textbf{rollout buffer} which stores the environment transitions collected with the current policy, and is emptied after performing the gradient update. Off-policy algorithms instead use an \textbf{experience replay buffer}, which stores older transitions, replacing the oldest ones once it reaches maximum capacity.

\subsection{Model-free or Model-based}

This division relies on whether or not the learning agent has access to a model of the environment $T(s, a)$. In \textbf{Model-free approaches} like DQN, PPO or DDPG, the agent learns in a true trial-and-error fashion -- 
it has no way of ``knowing" the consequences of an action until it tries it, and observes the outcome. 
On the other hand, \textbf{Model-based approaches} additionally learn a model of the environment, allowing the algorithm to do something akin to traditional planning algorithms by considering potential future states and actions, without actually having to execute them in the environment. This is famously present in the AlphaZero~\cite{silver_mastering_2017} algorithm that achieved superhuman performance in the game of Go, where one of the components is the Model-based Monte Carlo Tree Search (MCTS)~\cite{coulom_efficient_2006}. While Model-based approaches can provide an advantage in planning terms, the effectiveness of the agent will be limited by the quality of the learned model, which can be negatively affected if the environment is very complex, which is often the case in character animation. This is not the case with Model-free approaches, which do not require an accurate characterisation of the environment to be effective, although they lack the ability to explicitly foresee future states and actions. In this work, we focus on model-free algorithms due to their relevance to character animation.

\subsection{Single-agent or Multiagent}

Finally, an algorithm can be designed to work with either one agent, or multiple agents sharing the same environment. While most of RL development focuses on single-agent algorithms, those can be extended to become multiagent algorithms through Independent Learning (see Section~\ref{sec:independent-learning}). In competitive multiagent scenarios, algorithms typically use the concept of self-play, training against (possibly old) copies of themselves so that they can be robust when matched with a wide range of opponents. In cooperative scenarios, a common trend is introducing some type of centralization of information so that the agents can coordinate more effectively.

\subsection{Summary}

Looking at modern RL algorithms, it is difficult to cleanly separate them into different categories. Many of the most successful approaches combine different concepts, resulting in an algorithm that is, technically speaking, actor-critic and off-policy. That being said, if we are content with the definitions being fuzzy, we can still gain useful insights about the differences between them. 

Typically, value-based algorithms are also off-policy, and enjoy higher sample efficiency compared to policy-based ones. This is because any environment transition, once generated, can be used in perpetuity in multiple gradient updates. Conversely, policy-based algorithms like PPO make it possible to perform fewer gradient updates, because they involve optimizing the objective function directly through gradient ascent. This indicates that value-based methods can be better when it is difficult to obtain additional data, whereas policy-based methods can often be trained with smaller hardware needs, as they require fewer network updates.

%% file: text/04-single-algo.tex
\section{Single-agent RL Algorithms} \label{sec:single-algo}

In this section we provide descriptions of the most important modern RL algorithms. Due to the large quantity of different methods that appeared in the recent years, this is not meant to be a comprehensive list of all algorithms that could be applied in character animation, but rather the ones with the most relevance, either to this application in specific, or for the field in general. We also provide a sufficient amount of detail for the reader to grasp the main ideas of the algorithms, but refer them to the source papers for the remaining information. Specifically, we do not aim to include sufficient information that would make it possible to reimplement the algorithms without referring to the main paper or existing implementation, as that tends to be a very complex process, with many details being important.

\subsection{DQN} 

The first algorithm we discuss is \textbf{Deep Q Network (DQN)}~\cite{mnih_human-level_2015}, which gained prominence when it was used to master a suite of Atari games, achieving superhuman performance in some of them, drawing significant attention to the field. It is a prime example of a Value-based, Off-policy algorithm, and is remarkably simple in its basic form, allowing for a plethora of modifications which we discuss further in this section. DQN is a modern version of the older Q-Learning algorithm~\cite{watkins_q-learning_1992} which relies on the same principles, but only works on tabular domains (\ie with a finite number of states and actions).

In DQN, the agent is defined by a state-action value function $Q(s, a)$, represented with a neural network, which is then trained to approximate the real optimal Q function of the environment. This is achieved by performing gradient descent on a Bellman loss function, defined as 

\begin{align}
    \mathcal{L}_i(\theta_i) &= \EE_{s,a\sim\rho(\cdot)}\left[ (y_i - Q(s, a; \theta_i))^2  \right] \label{eq:dqn-grad} \\ 
    y_i &= \EE_{\substack{s,a\sim \rho(\cdot) \\ s'\sim T(s,a)}} \left[ R(s, a) + \gamma \max_{a'} Q(s', a'; \theta_{i-1}) \label{eq:dqn-grad1}  \right]
\end{align}
where $i$ is the current training iteration, $\theta_i$ are the weights of the neural network, and $\rho$ is the probability distribution over state-action sequences according to the behavior policy. Intuitively, the network is trained in a way similar to supervised learning, with the target being the empirical Q value of a given state-action pair, obtained by executing the policy and estimating the future utility using the same current Q function estimate. Typically, an automatic differentiation software is used to find the gradient of the loss function with respect to the network weights $\theta$, leading to the actual parameter update proportional to $\nabla_\theta \mathcal{L}(\theta)$

In order to ensure sufficient exploration, DQN uses $\epsilon$-greedy sampling. This means that given a value of $\epsilon \in [0,1]$, then while collecting data for optimization, the agent will choose a random action with a probability of $\epsilon$, and the optimal action (according to the current Q function estimate) with a probability of $1-\epsilon$. Commonly, $\epsilon$ is treated as a constant during a single training iteration, and progressively reduced to 0 as the training proceeds.

DQN also uses a replay buffer -- the collected data is stored and reused throughout the training, which is possible because DQN is an Off-policy algorithm. So the general flow of the algorithm is as follows. First, collect a batch of data using the current behavior policy (defined by the weights $\theta_i$ and some value of $\epsilon$), and add that data to the persistent replay buffer. Then, sample some data from the replay buffer, and perform gradient updates according to Equation~\ref{eq:dqn-grad}. Repeat this process, updating the weights and decreasing $\epsilon$ until convergence.

A crucial limitation of the DQN algorithm lies in the $\max$ operator of Equation~\ref{eq:dqn-grad1}. With a discrete action space, finding the optimal action is easy -- simply evaluate the function for each action, and then choose the best one. However, when dealing with continuous action spaces, this turns into a potentially non-trivial and nonlinear optimization problem, which in unfeasible to solve each time the agent needs to choose an action, which means that effectively, DQN is limited only to discrete action spaces. This can be avoided by changing the action space through discretization, or changing the algorithm (see Section~\ref{sec:ddpg}).

\subsection{Rainbow}~\label{sec:rainbow}

Over the last few years, many modifications of the core DQN algorithm have been developed, aiming at various improvements to its performance. Six of them were combined in the \textbf{Rainbow}~\cite{hessel_rainbow_2017} algorithm:

\begin{enumerate}
    \item Double Q-Learning~\cite{van_hasselt_deep_2015}
    \item Prioritized Experience Replay~\cite{schaul_prioritized_2016}
    \item Dueling Networks~\cite{wang_dueling_2016}
    \item Multi-step Learning~\cite{sutton_learning_1988}
    \item Distributional RL~\cite{bellemare_distributional_2017}
    \item Noisy Nets~\cite{fortunato_noisy_2017}
\end{enumerate}

The main ideas of them are as follows. \textbf{Double Q-Learning} trains two neural networks, decoupling the action selection from evaluation, in order to mitigate the problem of the learned Q networks overestimating the utilities. \textbf{Prioritized Experience Replay} changes the way in which old experience is sampled to optimize the Q network, so that more informative samples (\ie ones with large updates) occur more frequently. \textbf{Dueling Networks} have two computation streams, one for the value, and one for advantage, with some of the weights shared between them. \textbf{Multi-step Learning} involves a different way of bootstrapping the future rewards, by looking a few steps ahead (as opposed to just one). \textbf{Distributional RL} has the algorithm learn to predict the distribution of rewards, as opposed to just the mean reward itself. Finally, \textbf{Noisy Nets} improve exploration by using partially stochastic linear layers. For further details on each of these modifications, we refer the reader to the relevant papers.

Overall, Rainbow agents generally train faster and reach a higher performance than the baseline DQN agents. This comes at the cost of implementation complexity, with only some of the standard frameworks supporting it (see Section~\ref{sec:frameworks}), whereas DQN is very common, and relatively easy to implement in its basic form even for beginners.

\subsection{REINFORCE}

Similarly to how DQN is the simplest Value-based algorithm, \textbf{REINFORCE}~\cite{williams_simple_1992, sutton_policy_1999} is the original Policy-based method that is used with neural networks as function approximators. In its simplest form, it is a direct implementation of the Policy Gradient Theorem (see Section~\ref{sec:policy-gradient}). It involves training a neural network to directly approximate the optimal stochastic policy $\pi\colon \State \to \Delta \Action$, so that the expected total reward is maximized. This process is performed in an On-policy manner, with a fundamentally simple basic training loop:

\begin{enumerate}
    \item Execute the policy and collect a batch of experience.
    \item Perform a single gradient update of the policy and discard the data. 
    \item Repeat (1) and (2) until convergence.
\end{enumerate}

REINFORCE can employ some improvements to a naive implementation of the Policy Gradient Theorem. Recall the general policy gradient estimate:

\begin{equation}
    \hat{g} = \frac{1}{|\mathcal{D}|} \sum_{\tau \in \mathcal{D}} \sum_t \nabla_\theta \log \pi_\theta(a_t|s_t) R(\tau)
\end{equation}

While the reward $R(\tau)$ is computed for the entire trajectory, it is reasonable that when considering the action at a step $t_0 > 0$, we disregard the rewards obtained before, \ie for $t < t_0$, since the action at $t_0$ could not have affected them. Furthermore, subtracting a state-dependent baseline from the reward does not change its expectation, which means we can use the \textbf{advantage} instead, as defined in Section~\ref{sec:rew-dis-adv}. This is useful as it decreases the variance of the gradient estimation, leading to a faster and more stable training procedure.

The policy trained by REINFORCE is stochastic, which means that it outputs a distribution over actions $\Delta \Action$ rather than a single action. During training, the agent samples an action from the distribution in accordance with the policy gradient theorem. During deployment, it might be desirable to use the deterministic optimal action (\ie $\max_{a\in\Action} \pi_\theta(\cdot|s)$) for improved stability and predictability of the agent. Typically, a stochastic policy with continuous actions is modeled by a Normal (or Multivariate Normal for multidimensional action spaces) distribution. The neural network then outputs the mean action $\mu$, and the variance $\sigma^2$ under the assumption that the individual components of the action vector are uncorrelated. Alternatively, a global, state-independent variance can be maintained in the model, and adjusted during the training. In the case of discrete actions, the policy uses a Categorical distribution, with the neural network outputs corresponding to their logits. Mixed action spaces are also possible, and can be modeled as joint distributions.

REINFORCE, as well as the algorithms based on it, can be trained as Actor-Critic algorithms. The Actor is the policy network $\pi_\theta$ which is responsible for the actual decision making, while the Critic $V_\theta$ is trained using regular supervised learning techniques, and is responsible for the value estimation in computing the advantage.


\subsection{TRPO}

\textbf{Trust Region Policy Optimization (TRPO)}~\cite{schulman_trust_2015} is based on REINFORCE combined with the notion of a Natural Policy Gradient~\cite{kakade_natural_2001}. It aims to improve the amount of utility that the agent can obtain from a single batch of data. Recall that REINFORCE can only perform a single gradient update with a batch of data, usually with a constant or decaying learning rate. If the learning rate is too large, a small change in the policy weights can have a large impact on the behavior of the agent, making it difficult to tune while still maintaining good training efficiency.

In TRPO, there are several approximations that deviate from the theoretically-justified REINFORCE algorithm, but instead enable better practical performance. The key idea is the \textbf{trust region}, which corresponds to a constraint on the allowed KL divergence between policies in consecutive training steps. The general (theoretical) TRPO update in the training step $k+1$ is:

\begin{align}
    \theta_{k+1} = &\argmax_\theta \mathcal{L}(\theta_k, \theta) \label{eq:trpo-1}\\
    &\text{s.t. } \bar{D}_{KL}(\theta||\theta_k) < \delta \label{eq:trpo-2}
\end{align}
where $\delta$ is a hyperparameter defining the size of the trust region, and $\mathcal{L}$ is the surrogate advantage:

\begin{equation}
    \mathcal{L}(\theta_k, \theta) = \EE_{s, a\sim \pi_{\theta_k}} \left[ \frac{\pi_\theta(a|s)}{\pi_{\theta_k}(a|s)} A(s, a) \right] 
\end{equation}
which measures how the new policy performs compared to the old one. The most important feature of this approach is that theoretically, the KL divergence constraint ensures monotonic improvements with a sufficiently small $\delta$, while still being more sample-efficient than REINFORCE.  


Due to the $\argmax$ operator in Equation~\ref{eq:trpo-1}, each step is a constrained optimization problem, which is infeasible to solve hundreds or thousands of times throughout the training. For this reason, the actual algorithm uses additional approximations, resulting in an efficient, but complex Policy Gradient method. Due to this complexity, as well as the fact that other methods can be applied on minibatches of data and are more efficient (see: PPO, Section \ref{sec:ppo}), TRPO is rarely used in practice.

\subsection{PPO}\label{sec:ppo}

\textbf{Proximal Policy Optimization (PPO)}~\cite{schulman_proximal_2017} is the successor to TRPO, which through additional simplifications and approximations achieves comparable performance, but with a significantly simpler implementation. It is the de facto standard Policy Gradient algorithm at the moment, and is supported by all major libraries. 

Its core idea is to take several gradient update steps with an importance sampling term, without making the policy deviate too far from the original behavior policy. There are two main variants of PPO: PPO-Clip and PPO-Penalty. The former introduces a clipping term to the relative action probabilities in order to disincentivize large policy changes, as measured by KL divergence. The latter adds a penalty term to the loss function for the same effect. In practice, the PPO-Clip variant is more commonly used. Their respective loss functions are as follows:

\begin{align}
    L^{CLIP}(\theta) &= \EE \left[ \min (r_t(\theta)A_t, clip(r_t(\theta), 1-\epsilon, 1+\epsilon)A_t)    \right] \label{eq:ppo-clip} \\
    L^{KLPEN}(\theta) &= \EE \left[  r_t(\theta) A_t - \beta \text{KL}[\pi_{\theta_{old}}(\cdot|s_t), \pi_\theta(\cdot|s_t)] \label{eq:ppo-penalty} \right]
\end{align}
where $\epsilon$ is a hyperparameter, $r_t(\theta) = \frac{\pi_\theta(a_t|s_t)}{\pi_{\theta_{old}}(a_t|s_t)}$ is the probability ratio of the action, and $\beta$ is a coefficient which is adaptively adjusted during the training (if using the Penalty variant).

PPO typically uses an entropy bonus to improve exploration. This means that there is an additional term in the loss function proportional to the entropy of the policy $\pi_\theta$, resulting in the policy maintaining some randomness, even at the cost of efficiency. 

PPO is an Actor-Critic algorithm, with the Critic being responsible for value estimation that is then used to compute the advantages. The Critic network is typically trained by performing gradient descent on a Mean Square Error loss function between its outputs, and the empirical returns observed in the collected data. 

While PPO is typically considered as an On-Policy algorithm, that is not entirely accurate. A single PPO update typically involves several gradient updates, often performed on minibatches of experience, after which the data is discarded as is the case in REINFORCE. This means that while the data can be reused, it can only be done in a very limited way, unlike typical Off-Policy algorithms.

It is worth noting that policy-gradient algorithms (REINFORCE, TRPO, PPO), tend to be sensitive to the implementation details which we omit from this survey. This phenomenon is analyzed in three large-scale studies, to which we refer interested readers: \cite{henderson_deep_2017, engstrom_implementation_2020, andrychowicz_what_2020}.

\subsection{A3C, A2C}

The \textbf{Asynchronous Advantage Actor Critic (A3C)}~\cite{mnih_asynchronous_2016} algorithm, and its synchronous equivalent \textbf{Advantage Actor Critic (A2C)}~\cite{wu_scalable_2017} are largely of historical value now. The key idea of A3C is using multiple parallel copies of the environment, from which the data can be collected asynchronously, without needing to synchronize them between episodes, or between individual steps. This is meant to improve training efficiency by eliminating the time when an individual worker has to wait for the main process to collect their experience and perform a gradient update. 

When researchers continued working with A3C, they discovered that the asynchrony was not a necessary component, but rather an implementation detail, so they developed a simplified, synchronous version named A2C. This algorithm, in its essence, is very similar to REINFORCE with specific details such as using multiple parallel copies of the environment, and using a learned baseline for advantage estimation (which is not the original intent of REINFORCE, but is nevertheless an option in it).

\subsection{GAE}

While it is not a Reinforcement Learning algorithm in the same sense that DQN and PPO are, \textbf{Generalized Advantage Estimation (GAE)}~\cite{schulman_high-dimensional_2018} is a method that can be applied to any algorithms which use the notion of advantage. It is heavily based on the concept of TD-lambda~\cite{sutton_learning_1988}, and can be seen as its extension using Advantages. In the simplest sense, given a trajectory with rewards $r_t$ and a value estimation at each step $V_t$, we define the \textbf{Monte Carlo} advantage as:

\begin{equation} \label{eq:gae-mc}
    A_t = \sum_{i} \gamma^i r_{t+i} - V_t
\end{equation}
which is to say that we compute the expected total reward obtained by the agent, and subtract its estimated value. To use this expression directly, we need a full episode, which in certain environments might be infeasible or inefficient. Furthermore, as the sum of rewards depends on many decisions that the agent has yet to take in the future, the variance of this advantage estimation tends to be very large.

An alternative way is using \textbf{Temporal Difference (TD)} estimation by bootstrapping the expected returns, using the value function itself. Like before, given the rewards $r_t$ and value estimations $V_t$, we define the TD advantage, or one-step advantage, as:

\begin{equation} \label{eq:gae-td}
    A_t^{(1)} = r_t + \gamma V_{t+1} - V_t
\end{equation}

With an unbiased value estimator, the expected value of this expression is the same as Equation~\ref{eq:gae-mc}. At the same time, the variance can be significantly lower due to the lack of direct dependence on future rewards. With a biased value estimate, this becomes an example of the classic bias-variance trade-off, prevalent in Machine Learning.

Notice that intermediate n-step advantages can be defined by simply delaying the bootstrapping:

\begin{align}
    A_t^{(2)} &= r_t + \gamma r_{t+1} + \gamma^2 V_{t+2} - V_t \\
    A_t^{(n)} &= \sum_{i=0}^{n-1} \left[\gamma^i r_{t+i}\right] + \gamma^n V_{t+n} - V_t
\end{align}
which introduces a wide range of possible advantage estimation methods. What GAE proposes is using all n-step advantage estimates, weighted exponentially with a factor of $\lambda \in [0, 1]$:

\begin{equation}
    A^{GAE}_t = (1-\lambda) (A_t^1 + \lambda A_t^2 + \lambda^2 A_t^3 + \dots)
\end{equation}

This turns out to have a simple analytic expression that can be computed with a single pass algorithm. Empirically, GAE often noticeably improves the performance of RL algorithms, and is the de facto standard for advantage estimation in Actor-Critic algorithms.

\subsection{DDPG}\label{sec:ddpg}

An algorithm on the boundary between Value-based and Policy-based methods is the \textbf{Deep Deterministic Policy Gradient (DDPG)}~\cite{lillicrap_continuous_2015}. It is based on the notion of a Deterministic Policy Gradient~\cite{silver_deterministic_2014}, which is the gradient of a state-action value function with respect to the action. It can also be seen as an adaptation of the DQN algorithm to continuous action spaces.

In DDPG, we train two separate networks -- a state-action value network $Q_\phi\colon \State \times \Action \to \RR$, and a (deterministic) policy network $\pi_\theta\colon \State \to \Action$. The value network is trained in a way similar to DQN, with some tricks such as using a replay buffer and a target network to stabilize the training. The key difference lies in the $\max$ operator of Equation~\ref{eq:dqn-grad1}, which is not trivial with a continuous action space. This is where we use the second, policy network, trained to predict the optimal action according to the reward function. The Q network is optimized to minimize the following loss functions:

\begin{equation}\label{eq:ddpg-q-1}
    L(\phi) = \EE \left[ \left( Q_\phi(s, a) - y(r, s', d)  \right)^2 \right]
\end{equation}

\begin{equation}\label{eq:ddpg-q-2}
    y(r, s', d) = \left( r + \gamma (1-d) \max_{a'} Q_\phi(s', a') \right)
\end{equation}
where $(s, a, r, s', d)$ are the transitions in the replay buffer, with $s, s'$ being the current and next state, $a$ the action that was taken, $r$ the reward, and $d$ is equal to 1 if the state was terminal, and 0 otherwise. Then, the policy is optimized using gradient ascent to maximize the following objective:

\begin{equation}\label{eq:ddpg-pi}
    L(\theta) = \EE \left[ Q_\phi(s, \mu_\theta(s)) \right]
\end{equation}
This can then be differentiated using the chain rule, giving the policy gradient of:

\begin{equation}
    \nabla_\theta L(\theta) = (\nabla_a Q_\phi(s, a)) \cdot (\nabla_\theta \mu_\theta(s))
\end{equation}

Overall, DDPG can be seen as the simplest way of adapting DQN to continuous action spaces, without having to discretize the action space. Because it is off-policy, it can be more sample-efficient than competing on-policy algorithms, making it suitable for environments in which it is difficult to collect large amounts of data. However, its asymptotic performance is often worse than that of competing on-policy algorithms like PPO, which leads to its limited practical use in character animation.

\subsection{TD3}

\textbf{Twin Delayed DDPG (TD3)}~\cite{fujimoto_addressing_2018} is to DDPG what Rainbow is to DQN -- it introduces a series of tricks that significantly improve the algorithm's performance. The main changes are as follows:

\begin{enumerate}
    \item Clipped Double Q-Learning
    \item Delayed Policy Updates
    \item Target Policy Smoothing
\end{enumerate}

\textbf{Clipped Double Q-Learning} works similarly to Double Q-Learning described in Rainbow (Section~\ref{sec:rainbow}, using the smaller value of the two networks' outputs to prevent value overestimation. \textbf{Delayed Policy Updates} involves performing policy updates less frequently than Q function updates. Finally, with \textbf{Target Policy Smoothing}, noise is added to the target action, so that it is more difficult for the policy network to exploit errors in the Q function.

\subsection{SAC}

\textbf{Soft Actor-Critic (SAC)}~\cite{haarnoja_soft_2018} is in many ways similar to TD3, in that it is a modification of DDPG with certain changes introduced in order to improve its performance. Primarily, it uses entropy regularization by adding a term proportional to the policy's entropy to its optimization objective. This encourages the policy to remain stochastic, increasing exploration. Similarly to TD3, it uses Clipped Double Q-Learning, minimizing the Bellman loss of DDPG. However, there is no explicit policy smoothing, as SAC trains a stochastic policy instead of a deterministic one. As a result, the additional regularization is unnecessary, since actions are sampled from a nontrivial distribution.

Specifically, SAC learns three functions in parallel: the policy $\pi_\theta$, and two Q functions $Q_{\phi_1}$, $Q_{\phi_2}$, with the usual double Q-learning approach. Since they are trained on an entropy-regularized objective, the Q function optimization objective takes the following form:

\begin{equation}
    L(\phi) = \EE \left[ \left( Q_{\phi_i}(s,a) - y(r, s', d) \right)^2 \right]
\end{equation}

\begin{equation}
    y(r,s',d) = r + \gamma (1-d) \left( \min_{j=1,2} Q_{\phi_{j}} (s', \tilde{a}') - \alpha \log \pi_\theta(\tilde{a}'|s') \right)
\end{equation}
where $\tilde{a}' \sim \pi_\theta(s')$, and $\alpha > 0$ is the entropy regularization coefficient. Notice the similarity to Equations \ref{eq:ddpg-q-1} and \ref{eq:ddpg-q-2} of DDPG, with the key difference being that the objective now has a term proportional to the entropy of the action distribution $\alpha \log \pi_\theta(\tilde{a}'|s')$, and the action used for computing the Q value of the following step is taken directly from the behavior policy. 

When it comes to policy learning, as SAC learns a stochastic policy, it must output a distribution over the action space. The optimization takes the following form:

\begin{equation}
    L(\theta) = \EE \left[ \min_{j=1,2} Q_{\phi_j}(s, \tilde{a}') - \alpha \log \pi_\theta (\tilde{a}'|s) \right]
\end{equation}
where $\tilde{a}' \sim \pi_\theta(s')$. Notice again the similarity to Equation \ref{eq:ddpg-pi}, which confirms that SAC is, in its essence, an updated and improved version of DDPG.

It is important to keep in mind that while this is a general outline of the algorithm, there are many details that can significantly affect its performance. For more information on this, we refer the reader to the original paper, as well as the existing open-source implementations (Section \ref{sec:frameworks-algos}).

\subsection{Learning from Data}

As a general rule, Reinforcement Learning does not need expert data to train agents, instead using an environment that the agent can interact with. In some cases, however, it may be beneficial to use expert data to augment the learning process, or even eliminate the use of a simulation whatsoever. This is often referred to as \textbf{Imitation Learning}, because the agent learns to imitate the actions of an expert whose experience is shown to it.


\textbf{Behavior Cloning (BC)}~\cite{bain_framework_1999, ross_reduction_2011, daftry_learning_2016} is the simplest way to perform Imitation Learning. Its core idea is to treat Imitation Learning as a supervised learning problem, which given a dataset consisting of observations and actions, learns to map the former to the latter by training a classifier or a regressor. 

By including a model training phase in which the agent can interact with the environment, we can remove the requirement that the dataset contains the actions~\cite{torabi_behavioral_2018}. This significantly simplifies the data required to perform imitation learning, and enables learning by simply observing someone, much like humans do in the real world. However, the quality of the resulting behaviors is typically lower due to the fact that the dynamics model is only trained with on-policy data, which means that out-of-distribution errors are likely to occur. For this reason, if the data about actions is available, it is better to use it instead of relying only on observations.

The \textbf{Generative Adversarial Imitation Learning (GAIL)}~\cite{ho_generative_2016} algorithm represents the main alternative to Behavior Cloning. It relies on the concept of \textbf{Inverse Reinforcement Learning (IRL)}~\cite{ziebart_maximum_2008}, which means learning the reward function from demonstrations (as opposed to regular RL, where the agent learns a policy, or generates demonstrations, given the reward function). This, combined with the notion of adversarial learning known from \textbf{Generative Adversarial Networks (GAN)}~\cite{goodfellow_generative_2020}, and a PG-based update rule (originally TRPO) produces an efficient algorithm for Imitation Learning.

A common practice is using Imitation Learning methods in conjunction with standard, reward-based RL algorithms~\cite{goecks_integrating_2020}. This can be done by including a term derived from Imitation Learning either in the reward function, indirectly encouraging the agent to act similarly to the data, or by including it directly in the optimization objective.

\subsection{Summary}

We described the most noteworthy RL algorithms used in single-agent environments. From a practical point of view, we recommend either using the on-policy \textbf{PPO} with GAE for advantage estimation, or the off-policy \textbf{SAC}, which are the most popular algorithms of their respective categories. If the training data is difficult to obtain, SAC is typically better as it can reuse the data enabling higher sample efficiency. On the other hand, PPO often offers faster training in terms of the wall time by using parallelism in data collection and larger performance improvements per gradient update. If working with discrete actions, Rainbow or another version of DQN is also a viable choice. Finally, if one wants to incorporate real data in the training process, both BC and GAIL are strong options and can be integrated with other algorithms.

%% file: text/05-multi-algo.tex
\section{Multiagent RL Algorithms} \label{sec:multi-algo}

Here, we describe the algorithms that are adapted specifically for multiagent environments. Those are typically based on existing single-agent algorithms, with modifications that improve the training process by abusing the specific multiagent structure of the problem.


\subsection{Independent Learning} \label{sec:independent-learning}

Any single-agent algorithm can be used in a multiagent scenario by using \textbf{Independent Learning}, with the resulting algorithms typically called \textbf{IPPO}, \textbf{IDDPG} etc. This entails treating the other agents as parts of the environment, possibly including information about them in the observation, and then simply training as if it were a single-agent task. A simple way to accelerate this training process when all agents are identical is treating them as \textbf{homogeneous}, also called \textbf{Parameter Sharing}~\cite{terry_revisiting_2020}. With this approach, every agent receives their own observation and takes their own action, but they share the underlying neural network, and their experience can be combined for the training. Otherwise, if each agent has its own separately trained neural network, it is referred to as \textbf{heterogeneous}. It is possible to introduce some degree of heterogeneity by including an agent indicator in the agent's observations, as shown by Gupta et al.~\cite{gupta_cooperative_2017}.

\subsection{MADDPG}

\textbf{MultiAgent DDPG (MADDPG)}~\cite{lowe_multi-agent_2020} is an extension of the DDPG algorithm to explicitly use the structure of multiagent environments in the training procedure. It relies on the idea of \textbf{Centralized Training, Decentralized Execution (CDTE)}, which means that the algorithm can use global or hidden information, as long as the resulting agent only needs access to its own observations.

In multiagent environments, there are two main pieces of information that is not available during execution -- other agents' observations (or the global state), and the actions they take. However, when training in a simulation that we have total control over, these quantities are readily available, and so can be used in an Actor-Critic paradigm to optimize the Critic. Then, in the execution phase, only the agent's local observation is necessary for the Actor network to choose the action.

\subsection{MAPPO}


\textbf{MultiAgent PPO (MAPPO)}~\cite{yu_surprising_2021} is the result of extending the PPO algorithm analogously to the difference between DDPG and MADDPG. Because PPO is an Actor-Critic algorithm, the Critic similarly use centralized information such as other agents' observations and actions, while only the Actor is actually involved in the decision making during evaluation.

Since the concept and the name of MAPPO is generic, there are other works that introduce a similar extension~\cite{guo_joint_2020, liu_reinforcement_2020, hu_noisy-mappo_2021}. The details are different between those papers, but in the most robustly evaluated version of it uses the following five tricks:

\begin{enumerate}
    \item Value normalization through a running mean, for robustness with respect to the reward scale
    \item Value function input includes both global and agent-specific features, pruned to reduce the input dimensionality
    \item Data is not split into minibatches, and the algorithm uses relatively few training epochs
    \item The clipping factor is tuned as a trade-off between training stability and fast convergence
    \item Using death masking (inputs for dead or deactivated agents) through zero states with agent ID
\end{enumerate}

The resulting algorithm delivers results comparable with more sophisticated off-policy algorithms, while being viable to train using a single machine with one GPU. 




\subsection{QMIX}

\textbf{QMIX}~\cite{rashid_qmix_2018} and its derivatives are a family of algorithms that adapt Q-learning in cooperative scenarios, so that it can use centralized training, while maintaining the option to perform decentralized execution. The core idea is that the joint state-action value is a monotonic function of the state-action values of each individual agent. 

Consider the two extremes in terms of centralizing Q value estimation. On one hand, we have fully independent Q learning, where each agent optimizes their own reward, which can be a viable option as described in Section~\ref{sec:independent-learning}. On the other hand, we can also consider fully centralized Q learning, with a single network processing all agents' observations, and outputting their joint action. A simple middle ground can be found in Value Decomposition Networks (VDN)~\cite{sunehag_value-decomposition_2017}, where a joint Q function is expressed as a simple sum of the agent's individual Q functions:

\begin{equation}
    Q_{tot}(s, a) = \sum_i Q_i(s^i, a^i)
\end{equation}

QMIX introduces additional flexibility to this approach. It replaces the summation operator with an arbitrary function of the individual values, with the only restriction being that it is monotonic with respect to all its inputs:

\begin{equation}
    \frac{\partial Q_{tot}}{\partial Q_i} \geq 0, \forall i \in \mathcal{I}
\end{equation}

This is obtained by using a \textbf{mixing network} to represent $Q_{tot}$. The weights of the mixing network are the outputs of a set of hypernetwork~\cite{ha_hypernetworks_2016} conditioned on the environment state. This whole setup can be trained with significant information sharing between the cooperating agents, while in the execution phase, each agent only requires its own Q function $Q_i$.

Due to the popularity and effectiveness of QMIX, researchers have developed various modifications aimed at improving its performance even further~\cite{zhou_learning_2020, wang_qplex_2021, yang_qatten_2020, rashid_weighted_2020, son_qtran_2019}. However, recent work suggests that using regular QMIX with appropriate implementation details is enough to achieve results comparable or even superior to the more complicated algorithms~\cite{hu_riit_2021}.

\subsection{Summary}

When working with multiagent problems (\eg crowd simulation), we typically recommend using one of the single-agent algorithms and applying it with an \textbf{Independent Learning} approach as a starting point, with either \textbf{IPPO} or \textbf{ISAC} following the notation from Section~\ref{sec:independent-learning}, as well as \textbf{Parameter Sharing}. This is significantly simpler in implementation than using algorithms that introduce centralized communication, and can often yield competitive results. While adding some additional communication or centralization may be beneficial, MADDPG tends to be difficult to train in new environments.

%% file: text/06-single-appli.tex
\section{Skeletal Animation} \label{sec:applications}

Individual characters can be animated using kinematic or physics-based methods.  For the former case, the action space directly consists of kinematic poses or existing motion clips, and are defined based on motion capture data.
In contrast, physics-based methods have action spaces that directly or indirectly produce joint torques that drive the motion.
In this section, we first provide an abridged overview of RL as applied to kinematic methods.
We then shift our focus to physics-based methods. This begins with a general summary of the many nuances involved when using RL to control physics-based character movement, given that 
the default motions produced by RL algorithms for humanoid characters in the RL literature
are usually of low quality as compared to what is needed for computer animation applications.
We then categorize and review many of the recent methods and results for RL-based physics-driven
character animation.

\subsection{RL for Kinematic Motion Synthesis}


RL has a long-standing history of being used to learn kinematic controllers from motion capture data.
Here we provide a brief overview of work in this direction.
Motion generation can be framed as an RL problem where actions correspond to the choice of motion clips, as first applied to automatically-constructed graphs~\cite{arikan2002interactive,kovar2002motion,lee2002interactive}
and then in ways that were better tailored to locomotion tasks,
e.g.,~\cite{lee2006precomputing,treuille2007near}.  
Lee et al.~\cite{lee_motion_2010} later introduced the concept of continuous motion fields in support of a data-driven state-dynamics model. Optimal actions on this model are then learned via a table-based representations for the policy and value function. Modern motion matching methods can be seen as a short-horizon version of motion-fields. Ling et al. ~\cite{Ling2020} learn a latent action space using autoregressive variational autoencoders to define character controllers and thereby enabling optimal goal-based animations.


\subsection{The Many Challenges Beyond the Choice of Algorithm}

A considerable amount of thought is typically required to define a character movement task, articularly in a physics-based setting.
This begins with the design of the character, which involves making decisions related to joint torque limits, contact friction, mass distribution, joint limits, joint damping, simulation and control time steps, and more. The choice of action space can also have implications for the learned results. 
Available options include joint torques, joint PD-target angles, joint PD-target angle offsets from an available reference motion,  muscle-based activations, or more abstract actions for hierarchical control approaches. 
It is also sometimes possible to learn simplified actions spaces that avoid redundancies
or that sample from a reduced-dimensionality action manifold, which can possibly be learned as well.
The definition of the state of a character that is provided to the policy
can also have a significant impact. The pose can be represented as Cartesian joint locations,
or in a more traditional form consisting of a root position and orientation, followed
by a set of internal joint angles. Contact information can also be an important part of the state.

Next, the task rewards need to be designed, which may need to balance generic and possibly temporally-sparse
rewards related to the goals, rewards that encourage energy-efficient behavior, and shaping rewards
that help guide the solution in what can otherwise be an exceedingly-large search space.
Rewards also tend to work better when mapped to a fixed range, as commonly done using a negative exponential.
Episode termination criteria are also important, as they effectively constrain the search space
and, by virtue of providing no further rewards, also provide an implicit negative task reward.
Reward terms can be combined, using a weighted addition, e.g.,~\cite{peng_deeploco_2017} or in a multiplicative fashion, e.g.,~\cite{park2019learning}, and these 
choices can strongly impact the final learned policies.

The optimization criteria to define a natural human or animal motion are difficult to determine,
and thus a natural alternative is to instead seek to imitate motion capture data, either
as individual motion sequences, or as distributions using adversarial approaches. 
The choice of initial states for a task is important, as it can affect the task difficulty~\cite{reda2020learning}, 
and can also simplify the learning, as in the case of a motion imitation task where the 
initial states can be drawn from the given reference trajectory~\cite{peng_deepmimic_2018}.
Curriculum-driven learning can enable an easy-to-difficult learning order for a task~\cite{xie_allsteps_2020}.
Policies can be "warm-started" from existing solutions.  Prior knowledge should be used
where possible to set the relevant variances and exploration rewards.
External forces can also be allowed early on in the optimization~\cite{yu_learning_2018}, and then slowly withdrawn.
Hierarchical learning can also be leveraged, by first learning low-level control 
that operates at a fine time scale, followed by higher-level control that allows
for long-horizon tasks~\cite{peng_deeploco_2017}.

The algorithms themselves are challenging to work with, with a typical improve-and-test debugging iteration 
requiring between hours and days, depending on the task difficulty and the availability of compute.
In many cases, wall-clock time is a more important consideration than sample-complexity, 
and algorithms whose common implementations support a high-degree of parallelization, e.g., PPO, are then sometimes preferred over that are more difficult to parallelize, e.g., SAC.
Tuning the algorithm hyper-parameters plays an important role in the learning efficiency and success,
and may require grid search or other hyper-parameter optimizations.
The results of model-based trajectory optimization can be used to guide policies towards suitable solutions
for difficult tasks. Debugging RL tasks is also an important skill, and points to initially
working with simplified or more-constrained systems, visualizing reward terms, understanding the
limitations of physics engines, and much more.
More specific algorithmic features to consider include the use of a Huber loss instead of the conventional
quadratic loss for Q-learning, considering various forms of conservative Q-learning, choice of temporal-difference
horizon, and more.

Many simulated robotic control environments are standard benchmarks for RL algorithms. MuJoCo~\cite{todorov_mujoco_2012} and PyBullet~\cite{coumans_pybullet_2016}, two of the most commonly used physics simulation engines in RL, provide several robot models with a Gym~\cite{brockman_openai_2016} interface. These robots range from abstract ones like Hopper or Reacher, through animal-like Ant and HalfCheetah, to more human-like ones like Humanoid and Walker2d. While not realistic, they share many of the principles of skeletal character animation.

We next review work that uses reinforcement learning to develop a variety of full-body motion skills for physics-based characters.
These leverage many of the insights described above.

\subsection{RL for Individual Character Skills}

For the remainder of this section, we further categorize methods into: (a) those which use motion capture data,
typically as a key part of the imitation objective, and (b) methods that use a more general ``pure'' learning objective.
In both cases, there exists a variety of prior art that is entirely model-based or uses other optimization methods.
However, for our purposes here, we restrict ourselves to methods that use reinforcement learning for motion imitation. 


\subsubsection{Motion imitation RL methods}


One of the first RL methods to be able to successfully imitate motion capture data, including highly dynamic motions
such as flips, uses data from a stochastic planning method, first developed as an open-loop 
trajectory optimization method~\cite{liu2010sampling}. Building on this type of method,
the work of \cite{liu2016guided} proposed to use data from multiple runs of the stochastic trajectory optimizer
to then learn a state-conditioned feedback policy. The desired motion sequence is divided into a sequence of 0.1~s duration
control fragments, and for each such fragment computes a multivariate linear regression of the actions with respect to the state.  
This yields a simple linear policy for actions as a function of the state, for the duration of the control fragment.
This model is then able to robustly imitate walking, running, spin-kicks, and flips, as well as transitions.
Further work has then shown how learned control fragments can be treated as abstract actions, which can be resequenced 
using deep Q-learning~\cite{liu2017learning}, and can further be adapted to learn basketball playing skills~\cite{liu2018learning}.

\begin{figure}[htb]
  \centering
  \includegraphics[width=1\linewidth]{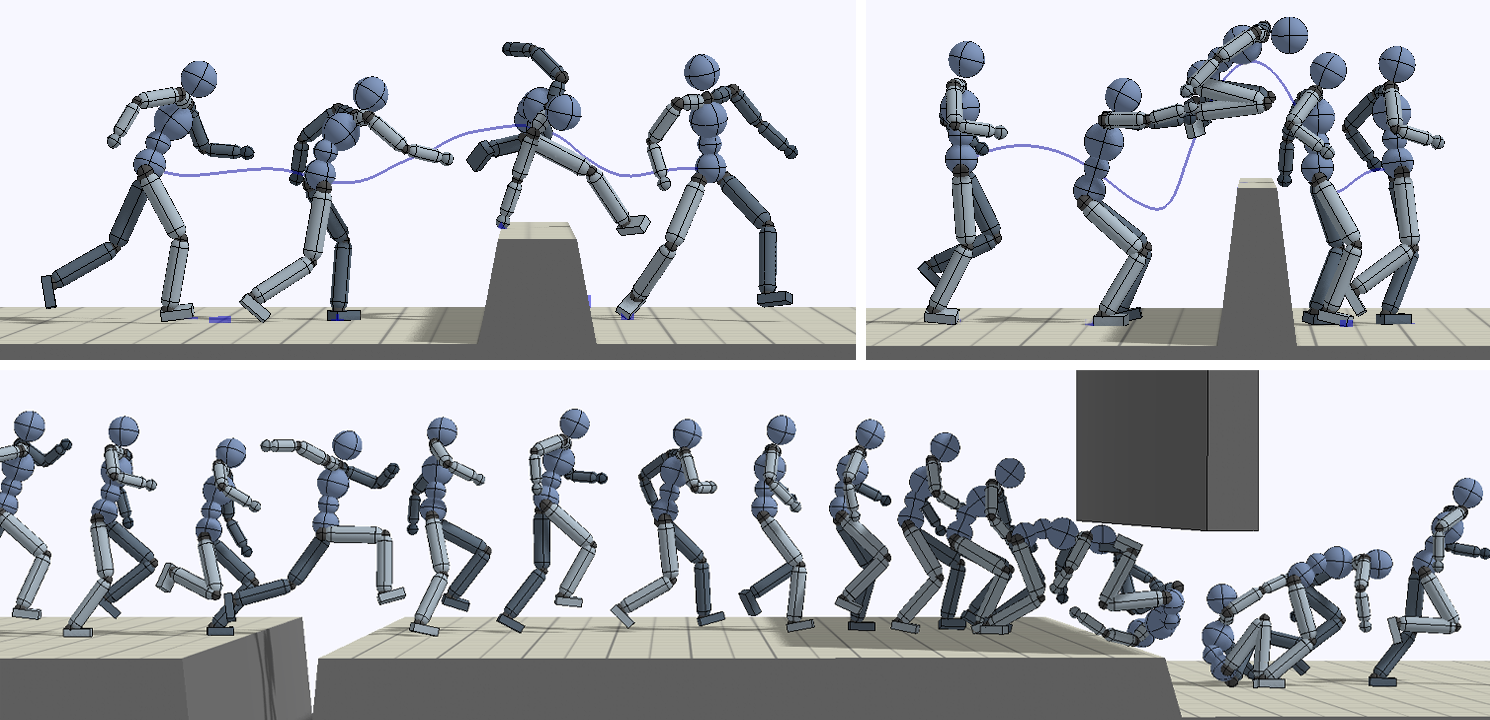}
  \caption{ 
  Imitation-based Learning. Proposed methods as in~\cite{peng_deepmimic_2018} allow to successfully synthesize animations from motion capture data. In other works, as in~\cite{peng_amp_2021}, they combine such techniques with the possibility of adding low-level behaviours to control the production of high-complexity animations.}
  \label{fig:imitation-learning}
\end{figure}

The use of policy gradient RL methods to imitate human motion capture clips was first explored by
Peng et al.~\cite{peng_deeploco_2017} for a variety of walking gaits.  This also introduced a hierarchical reinforcement learning approach, 
with a low-level policy first being trained to reach target stepping locations while also striving to imitate the reference motion.
A high-level policy then operates once for every walking step, generating step targets in support of tasks,
including control of a ball with the feet, navigating paths, and avoiding dynamic obstacles.
Peng et al.~\cite{peng_deepmimic_2018} further develop the imitation learning approach 
to train controller for a diverse set of motions, including highly-dynamic spin kicks and flips for humanoids, 
sequencing such motions, and using the same imitation approach for quadruped controllers.
Imitation-based learning of a wider variety of quadruped gaits, including sharp turns, is demonstrated in~\cite{peng2020learning},
along with successful transfer to quadruped robots. 
Peng et al.~\cite{peng_amp_2021} use ideas from adversarial imitation learning by combining a reward function to control the high-level behaviors, with low-level controls specified with an unstructured dataset of motion clips. This method can be used on both humanoid and non-humanoid models. It produces high-quality animations that match tracking-based methods, but the training process can still be prone to mode collapse, as is common in GAN-like algorithms. Some of these examples in Imitation Learning are shown in Figure~\ref{fig:imitation-learning}.
The choice of action space is also shown to have an impact the speed and quality of imitation-based learning~\cite{peng2017learning}.


Computer vision based pose tracking can also be used
as a source of motions to imitate, allowing robust control policies to be learned from video clips~\cite{peng2018sfv}.
Isogawa et al.~\cite{isogawa_optical_2020} construct an end-to-end pipeline that converts Non-Line-Of-Sight measurements to 3D human pose estimation by employing a diverse set of techniques, including an RL-based humanoid control policy. Yuan et al.~\cite{yuan_simpoe_2021} introduce the SimPoE framework, which trains an RL agent to control a physics-based character to estimate plausible human motion, while conditioning it on a monocular video. 

The majority of the works described above develop control policies that only reproduce
single clips, or a specific set of motion clips.  The motion to imitate plays a role via the reward,
but is not provided to the policy as an input. The policies are conditioned on a time or motion phase. 
An important next step has been to reproduce a richer variety of motions
by conditioning the policy on a short time window of the future motion to imitate.  This can also
be seen as a generalized form of learned inverse dynamics, with a longer anticipatory window as
needed to make motion corrections for more difficult motions.
Chentanez et al.~\cite{chentanez2018physics} first develop this type of conditioning
and apply it to large motion datasets. Significant further developments follow from improvements
that target scalability, motion transitions, motion quality, generalization, and learning efficiency~\cite{park2019learning,bergamin2019drecon,wang2020unicon,won2020scalable}. 
These methods are further extended to work with muscle-based actuations~\cite{lee2019scalable},
a large diversity of body shapes~\cite{won2019learning}, and producing large motion variations even 
from a single motion clip~\cite{lee2021learning}.
Other work shows how to allow for more flexible forms of imitation~\cite{ma2021learning}, 
and that leverage residual external forces to enable learning more challenging motions~\cite{yuan2020residual}.
Imitation-based controllers can also be used to learn a latent human-like action space 
via distillation (``neural probabilistic motor primitives''),
which can then be used as an abstract action space for new tasks~\cite{merel2020catch}.
Similarly, Luo et al.~\cite{luo_carl_2020} learn a natural action distributions from reference motions for quadrupeds,
while a GAN-based controller reproduces suitable actions based on user-input. This is followed by high-level DRL fine-tuning.

\subsubsection{Pure objective RL methods}

Reinforcement learning has also been successfully used for full-body character animation without an imitation objective.
Here, the objective can be framed in terms of rewards that include energy, progress towards a goal, stylistic hints,
and regularization terms.  


Model-predictive control (MPC) methods, which iteratively re-plan and then execute the first action, 
have been successfully employed for humanoid animation and are a form of model-based RL.
The work of Tassa et al.~\cite{tassa_synthesis_2012} demonstrated the online use of iLQG (Iterative Linear Quadratic Gaussian) 
trajectory optimization for online control of humanoid characters for a variety of tasks, including getting up, using a 0.5~s 
planning time horizon. Sampling-based methods can also be used to achieve trajectory optimization over a finite planning horizon,
and have been explored in detail by H{\"a}m{\"a}l{\"a}inen et al.~\cite{hamalainen2014online,hamalainen2015online}. 
Online trajectory optimization and policy learning can also be used in a mutually supportive fashion~\cite{rajamaki2017augmenting}, with the
policy serving to accelerate the trajectory optimization, and the trajectory optimization helping to bootstrap the policy learning. In addition, trajectory optimization can benefit from more complex search spaces, for instance by including contact points~\cite{Mordatch2012} to improve simultaneously both, trajectory and policy learning.

Actor-critic methods for RL can more easily tackle motion tasks, such as locomotion, 
by being provided with task-specific action abstractions.  For example, the action space can consist of
a discrete set of existing controllers, with a high-level actor-critic controller being trained to make a discrete
selection among the set of available controllers at each step of the gait. This setup is used with a $k$NN-based value function
approximator to achieve high-level objectives by Coros et al.~\cite{coros2009robust}.
An abstracted, tailored action space is used by Peng et al.~\cite{peng_dynamic_2015} to include a continuous action space 
as defined by a conveniently-parameterized finite-state machine controller.
A $k$NN-based actor-critic pair is then used to train dog-like and bipedal models to traverse variable terrain.
Later, Peng et al.~\cite{peng_terrain-adaptive_2016} 
develop a mixture-of-experts based Actor-Critic algorithm named MACE for improved performance on a similar dynamic locomotion task,
this time using deep neural networks for the actors and critics, and thereby eliminating some of the feature engineering required by the previous approach. 

\begin{figure}[htb]
  \centering
  \includegraphics[width=1\linewidth]{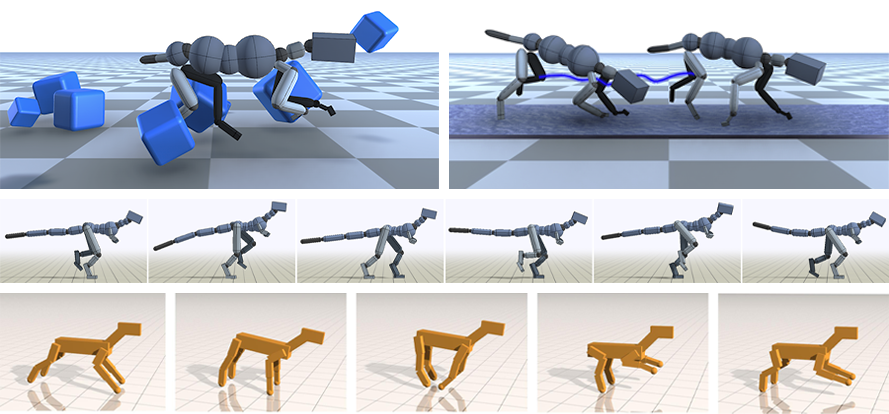}
  \caption{\label{fig:morphology} 
  There is a wide variety of methods that also address the synthesis of animations for quadrupedal or arbitrary morphology~\cite{luo_carl_2020}~\cite{peng_deepmimic_2018}~\cite{yu_learning_2018}. While the limited amount of motion capture data introduces an additional challenge, such methods try to overcome this constraint by covering a wide range of techniques, from imitation-based approaches to pure objective RL.}
\end{figure}

Can policy-gradient RL algorithms be used with pure learning objectives to generate natural human movement, as opposed to the unrealistic frenetic motions commonly seen resulting from popular RL benchmarks?
Yu et al.~\cite{yu_learning_2018} encourage symmetric and low-energy motions by appropriately modifying the loss function of the algorithm, by adding the so called mirror-symmetry Loss to the usual surrogate loss of PPO. This allows for high-quality motions without 
using any imitation of motion examples. This is particularly important for non-humanoid characters for which there is no motion data available. Example non-humanoid models that can be trained this way are in Figure~\ref{fig:morphology}.
Abdolhosseini et al.~\cite{abdolhosseini_learning_2019} further investigate multiple methods of incorporating symmetry constraints for skeletal animation tasks, inspired by the observation that human and animal gaits found in nature are typically symmetrical. They use the PPO algorithm with four options of enforcing symmetry on the learned policy. They show that this can in fact be harmful to the training process, but in the end can produce higher quality motions. 
Xie et al.~\cite{xie_allsteps_2020} explore a curriculum-based learning solution to train characters to walk and run over a wide range of stepping stones, with varying step heights, lengths, yaw angles, and step pitches, in the absence of 
an imitation objective. PPO is used in conjunction with a parameterized generator of individual steps, and the learning curricula
advance the step difficulty in several different ways. PPO is used to train the physical legged model. 


To further improve on the realism, more biomechanically accurate models can be considered. 
The NeurIPS conference hosted three challenges using the osim-rl platform (see Section~\ref{sec:frameworks-envs}): "Learning to Run" (2017), "AI for Prosthetics" (2018) and "Learn to Move - Walk Around" (2019)~\cite{kidzinski_learning_2018-1}, all of which dealt with different aspects of controlling a human body model. The leading solutions used learned models of the environment, and off-policy algorithms such as DDPG.
Jiang et al.~\cite{jiang_synthesis_2019} continue this research direction by using another human body model based on the OpenSim~\cite{seth_opensim_2011} platform. With muscles outnumbering joints, the larger action space of biomechanical models 
can be significantly more expensive to train.  This paper addresses this by allowing the optimization to nevertheless operate in joint 
actuation space, as afforded by two neural networks that model the state-dependent torque limits and the metabolic energy. The overall approach is agnostic to the choice of RL algorithm.


Non-locomotion tasks are also important for full-body character animation.
Kumar et al.~\cite{kumar_learning_2017} use an algorithm based on MACE for the task of teaching a virtual humanoid model to safely fall by minimizing the maximal impulse experienced by its body. They train a mixture of actor-critic networks associated with all possible contacting body parts, and further use a form of hierarchical reinforcement learning, with an abstract policy deciding the high-level behavior, and a joint policy responsible for actually executing the action. 
Clegg et al.~\cite{clegg_learning_2018} consider the problem of simulating the movement of a human dressing themselves using a combination of physics simulation and RL. They use a virtual human-like model and is tasked with putting on one of three different pieces of clothing. The process of getting dressed is divided into several subtasks, each of which is treated as a Reinforcement Learning problem with an appropriate reward function. Subtasks are trained using TRPO, and then sequenced together to produce a full motion. 
Yin et al.~\cite{yin_discovering_2021} explore the problem of learning diverse jumping motions, including high jumps. 
For a given takeoff state, a curriculum is used to learn a policy for increasing bar heights.
The space of takeoff states is then explored using Bayesian diversity search, to synthesize a diverse set of jumping styles,
including jump well-known techniques such as the Fosbury flop.

%% file: text/07-multi-appli.tex
\section{Crowd Animation}\label{sec:appli-crowd}

In this section we discuss the specific work that used RL algorithms in Crowd Animation scenarios, as well as the challenges that make this task distinct from single-agent cases.
Unlike Skeletal Animation, Crowd Animation typically uses multiagent algorithms so that each individual agent has access to its own information, but not necessarily to the global state. While the task can often be seen as fully cooperative (\eg making a realistic simulation), this enables more realistic behaviors, and not using a centralized controller enables easier scaling to different numbers of agents. We focus on applications pertaining to the challenge of multiple agents navigating in a shared environment, and omit the discussion of more advanced topics around coordination and division of tasks, considering them to be out of scope of this work.


\subsection{Challenges of Crowds}

The key factor distinguishing cooperative multiagent learning from single agent RL is the phenomenon of nonstationarity. Typically, RL algorithms assume that the environment is stationary, which means that the environment dynamics (represented by the transition function) remain the same throughout the training. That is not true in multiagent training as observed by a single agent -- as other agents learn, their policies change, which affects the perceived environment dynamics.

In the case of Crowd Simulation specifically, there is also the question on how exactly to represent the physics of the problem. While some works use holonomic cartesian controls in which each agent can move in any direction, this is not entirely realistics, and instead, polar controls may be used, where an agent decides its linear motion and turning left or right. Furthermore, the agents may either control their velocities directly, or apply accelerations to their motion. While these approaches can be seen as nearly equivalent, it is still necessary to choose one, which may impact the final performance in nontrivial ways.

In certain cases, competitive and general-sum scenarios may be relevant, which carries additional complications. Most notably, evaluation of trained systems is challenging in the absence of an expert model or an external performance measure. This is because the typical training paradigm relies on self-play, and a winrate against a copy of itself cannot be reliably translated to objective performance. Furthermore, the details of the self-play procedure can also impact the training. Finally, as training progresses, agents might learn to specialize to play against specific strategies, forgetting about their older versions, and underperforming when matched up against them.

Finally, it is worth mentioning that many of the challenges in Skeletal Animation, still apply here. Depending on the physical model of the agent behaviors and interactions, the exact choice of the actuator may be important for the learned policies. Similarly, designing the reward function is crucial for good performance -- a reward that is too sparse may be prohibitively difficult, whereas one that is adapted to be more dense, may lead to unexpected behaviors. Finally, it is often desirable to have agents exhibit human-like behavior, but this task in itself is not well-defined, and it may be helpful to use real-world data in order to generate a specific reward function.


\subsection{Applications}

Long et al.~\cite{long_towards_2018} apply an RL-based approach to the task of collision avoidance. While this is not exactly the same thing as character crowd animation, collision avoidance is nevertheless a significant component of crowd simulation systems. They train a policy which receives as input a depth map, the goal's relative position and the current velocity, on the task of reaching the goal and avoiding collisions with other agents in the shared environment. Because they also deploy them on real robots, there is additional emphasis on avoiding collisions, with contact between two agents leading to removing both of them from operation with a large negative reward. They train the policy using the PPO algorithm, and show that this method results in higher success rates and more efficient policies compared to ORCA.
With a similar approach, Lee et al.~\cite{lee_crowd_2018} apply the DDPG algorithm to the task of basic crowd simulation -- a number of agents in a shared environment move through it, and attempt to reach their respective destination. They use polar dynamics, with the RL agent setting a linear velocity and a rotation at each timestep. The agents receive a positive reward for getting closer to their goals, a negative reward when they collide, and there is also a regularizing reward term that encourages smooth movement. They show that this setup is sufficient to obtain agents that reach their goals and avoid collisions, although the results are still imperfect. There is also no regard for human-like behavior.

To explicitly combine ORCA with RL methods on the same standard crowd simulation task, Xu et al.~\cite{xu_local_2020} introduce the ORCA-DRL algorithm. They use PPO to predict a preferred velocity at any given step, which is then used as an input to ORCA. This is then responsible for actually avoiding collisions. They show that this approach leads to successful collision avoidance, which is not surprising given its reliance on a classical collision avoidance algorithm. It also does not consider whether the behavior is human-like.
Sun et al.~\cite{sun_crowd_2019} use a different approach -- they train four "leader" agents responsible for guiding parts of the crowd, while the remaining agents follow their respective leaders. They use PPO with a recurrent LSTM~\cite{hochreiter_long_1997} policy and combine it with a classical collision avoidance algorithm RVO. They train the agents to act in an unknown environment with dynamic obstacles. The resulting behaviors manage to achieve their goals, but do not take into consideration whether the behavior is human-like. This method also still relies on classical collision avoidance algorithms.


Haworth et al.~\cite{haworth_deep_2020} introduces a method on the borderline between single character and crowd animation. By employing a method based on the ideas of Hierarchical Reinforcement Learning, they train two policies interacting with one another. The high-level policy is responsible for navigation and reaching global goals. It sets objectives for the low-level policy which directly controls the joints of a humanoid model. They use this on multiple characters in a shared environment, with the low-level policies shared between them. Both policies are trained using PPO, with the low-level policy learning to match motions from a database of stepping actions using a PD controller.

H\"uttenrauch et al.~\cite{huttenrauch_deep_2019} introduce a method that, while not directly applied in crowd simulation, is nevertheless very relevant. Their work focuses on swarm systems, which are inherently similar to crowd scenarios -- they both focus on a large number of agents acting in a shared environment, often with a shared goal, with the individual agents typically being indistinguishable. They introduce a method called Mean Feature Embedding, similar to existing Relation Networks~\cite{santoro_simple_2017, zambaldi_relational_2018}. This approach uses a modified neural network architecture that ensures permutational invariability between identities of different neighboring agents that are perceived by another agent. This inductive bias can accelerate the training process, and improves scaling to different numbers of agents in the environment.
Alonso et al.~\cite{alonso_deep_2020} explore the applicability of RL methods for the task of crowd navigation in AAA games. They use a large and complex 3D environment built on Unity, modelled after real games that typically use a Navigaton Mesh (NavMesh)~\cite{snook_simplified_2000} approach. They use a recurrent LSTM network to give their agents memory, and use the SAC algorithm for policy optimization. As inputs, the agents receive their absolute positions, relative goal position, their speed and acceleration, as well as 3D occupancy maps obtained via box casts, and depth maps from ray casts. They show that this approach has a high level of success, and can enable more flexible map designs, without requiring the designers to specify each possible link.

Zou et al.~\cite{zou_understanding_2018} consider the problem of understanding and predicting crowd behavior specifically from the perspective of Imitation Learning. They introduce a new framework named Social-Aware Generative Adversarial Imitation Learning (SA-GAIL) which is trained to replicate behavior recorded in demonstrations, while disentangling the different factors of decision-making in pedestrian movement. This allows them to obtain a human-understandable interpretation for the model's predictions, as well as for the real data. They use the TRPO algorithm for policy optimization and show that this approach can produce high-quality, interpretable behaviors.
A different approach for using data to obtain more human-like behaviors is used by Xu and Karamouzas~\cite{xu_human-inspired_2021}. They use the concept of Knowledge Distillation introduced earlier by Hinton et al.~\cite{hinton_distilling_2015}. Along with the standard PPO algorithm for policy optimization, they train a neural network in a simple supervised way, mapping observations to actions based on data from a crowd motion dataset. Then, the outputs of this network are used as an additional source of reward for the policy learning, encouraging the agent to act similarly to what the supervised network predicts. This way, they obtained more human-like behaviors on typical crowd simulation scenarios, as compared to a regular RL baseline, without a detriment to their performance.

We summarize the algorithms from the papers listed in Sections \ref{sec:applications} and \ref{sec:appli-crowd} in Table \ref{tab:algorithms}. While there is a decent diversity of physics engines, as well as a split between TensorFlow in PyTorch for the neural network optimization, the RL algorithm of choice is predominantly PPO.

\input{tables/algorithms}

%% file: tables/algorithms.tex
\begin{table*}[ht]\centering
\caption{A summary of the DRL algorithms, simulation engines, and neural network frameworks in the described papers, where applicable and stated in the paper or the provided source code. $^{1}$ Value Iteration, $^{2}$ Open Dynamics Engine, $^{3}$ Temporal Difference learning, $^{4}$ Maximum A Posteriori Policy Optimization.}\label{tab:algorithms}
\begin{tabular}{ccccc}
Citation                           & Year & Algorithm    & Physics Simulation engine & NN Framework \\ \hline
\cite{peng_dynamic_2015}           & 2015 & VI$^{1}$     & Box2D                     & --           \\
\cite{peng_terrain-adaptive_2016}  & 2016 & MACE         & Bullet                    & Caffe        \\
\cite{liu2017learning}             & 2017 & DQN          & ODE$^{2}$                 & Theano       \\
\cite{peng_deeploco_2017}          & 2017 & TD$^{3}$     & --                        & --           \\
\cite{peng2017learning}            & 2017 & TD           & --                        & --           \\
\cite{kumar_learning_2017}         & 2017 & MACE         & DART                      & Caffe        \\
\cite{liu2018learning}             & 2018 & DDPG         & ODE                       & Theano       \\
\cite{peng_deepmimic_2018}         & 2018 & PPO          & Bullet                    & TensorFlow   \\
\cite{peng2018sfv}                 & 2018 & PPO          & Bullet                    & TensorFlow   \\
\cite{chentanez2018physics}        & 2018 & PPO          & MuJoCo                    & TensorFlow   \\
\cite{yu_learning_2018}            & 2018 & PPO          & DART                      & TensorFlow   \\
\cite{clegg_learning_2018}         & 2018 & TRPO         & DART                      & PyTorch      \\
\cite{long_towards_2018}           & 2018 & IPPO         & Stage                     & TensorFlow   \\
\cite{lee_crowd_2018}              & 2018 & IDDPG        & --                        & TensorFlow   \\
\cite{zou_understanding_2018}      & 2018 & TRPO + GAIL  & --                        & TensorFlow   \\
\cite{park2019learning}            & 2019 & PPO          & DART                      & TensorFlow   \\
\cite{bergamin2019drecon}          & 2019 & PPO          & Bullet                    & TensorFlow   \\
\cite{lee2019scalable}             & 2019 & PPO          & DART                      & PyTorch      \\
\cite{won2019learning}             & 2019 & PPO          & DART                      & TensorFlow   \\
\cite{abdolhosseini_learning_2019} & 2019 & PPO          & Bullet, MuJoCo            & PyTorch      \\
\cite{sun_crowd_2019}              & 2019 & IPPO + RVO   & Unity3D                   & --           \\
\cite{huttenrauch_deep_2019}       & 2019 & ITRPO        & --                        & TensorFlow   \\
\cite{peng2020learning}            & 2020 & PPO          & Bullet                    & TensorFlow   \\
\cite{wang2020unicon}              & 2020 & PPO          & Flex                      & -            \\
\cite{won2020scalable}             & 2020 & PPO          & Bullet                    & TensorFlow   \\
\cite{yuan2020residual}            & 2020 & PPO          & MuJoCo                    & PyTorch      \\
\cite{merel2020catch}              & 2020 & V-MPO$^{4}$  & MuJoCo                    & NumPy        \\
\cite{luo_carl_2020}               & 2020 & PPO          & Bullet                    & TensorFlow   \\
\cite{Ling2020}                    & 2020 & PPO          & Bullet                    & PyTorch      \\
\cite{isogawa_optical_2020}        & 2020 & PPO          & MuJoCo                    & TensorFlow   \\
\cite{xie_allsteps_2020}           & 2020 & PPO          & Bullet                    & PyTorch      \\
\cite{xu_local_2020}               & 2020 & IPPO + ORCA  & --                        & PyTorch      \\
\cite{haworth_deep_2020}           & 2020 & IPPO, MADDPG & Bullet                    & Caffe        \\
\cite{alonso_deep_2020}            & 2020 & SAC          & Unity3D                   & --           \\
\cite{peng_amp_2021}               & 2021 & PPO + GAIL   & Bullet                    & TensorFlow   \\
\cite{lee2021learning}             & 2021 & PPO          & DART                      & TensorFlow   \\
\cite{ma2021learning}              & 2021 & PPO          & Bullet                    & PyTorch      \\
\cite{yuan_simpoe_2021}            & 2021 & PPO          & MuJoCo                    & PyTorch      \\
\cite{yin_discovering_2021}        & 2021 & PPO          & Bullet                    & PyTorch      \\
\cite{xu_human-inspired_2021}      & 2021 & IPPO         & -                         & PyTorch     
\end{tabular}
\end{table*}

%% file: text/08-interaction.tex
\section{Human Interaction}\label{sec:interaction}

While not directly a part of Character Animation, interaction between humans and learning-based agents are highly relevant to its applications, notably for Virtual Reality games. For this reason, in this section we describe some of the work towards interactive RL agents. For agents to be interactive, it means that they must be capable of acting in a shared environment with a human-controlled agent, in such a way that they retain their performance on their original goals, while simultaneously reacting to the human's actions appropriately. This is far from trivial, especially when the human behavior significantly differs from the behavior of the trained agents. 

An important concept to discuss is the \textbf{Theory of Mind (ToM)}~\cite{premack_does_1978}. Stemming from developmental psychology, ToM refers to the ability that humans and some other animals possess, of reasoning about the internal state of someone else -- their goals and beliefs. As Rabinowitz et al.~\cite{rabinowitz_machine_2018} show, it is also possible to train RL agents so that they can learn ToM of other agents in their environment by observing their actions.

Chodhury et al.~\cite{choudhury_utility_2019} consider whether it is worthwhile for an agent to learn a full environment model, to learn a ToM model of the human, as opposed to using a model-free approach, in order to cooperate with a human agent. They use an autonomous driving task and show that general black-box model-based methods can work as well as ToM learning, and both of them outperformed the model-free approach.

Carroll et al.~\cite{carroll_utility_2019} analyze this problem in the general case of cooperative multiagent reinforcement learning, using an environment based on the game Overcook, which requires a high level of cooperation. They find that agents trained in the usual ways, such as with self-play or population-based training, perform significantly when paired with a human played, as compared to their original group performance. They introduce a method based on training a Behavior Cloning agent on data collected from human gameplay. The policy of this agent is then frozen, and it is used as part of the environment dynamics for the actual agent we want to train. Despite the BC agent's low quality, this turns out to be sufficient to improve the performance of the actual agent when evaluated together with a human player.

Christiano et al.~\cite{christiano_deep_2017} include humans in the loop in the training phase, as opposed to enabling cooperation with them. They introduce a method with which it is not necessary to specify a reward function for an agent to optimize. Instead, the algorithm produces demonstrations which are then judged by the human, allowing it to assign reward values from which it learns. This way, RL agents can learn complex behaviors which are not trivial to define mathematically, instead relying on human preferences, while only requiring human input on about 1\% of the actual frames used during training.

%% file: text/09-frameworks.tex
\section{Frameworks} \label{sec:frameworks}

\input{tables/frameworks}

In this section, we discuss the most relevant libraries and frameworks used for training RL agents, which are then applied to character animation. Because the field of Reinforcement Learning relies on neural networks, we begin by describing main frameworks used in Deep Learning. Those are responsible for efficiently performing algebraic operations on tensors (here understood as n-dimensional arrays), using parallelism when possible. They also take care of computing the gradients of functions with the backpropagation algorithm, enabling gradient-based optimization. They do that either on CPU or GPU, sometimes also supporting TPU (Tensor Processing Unit). Then we move on to libraries that function as backbones for RL tasks, by providing standard implementations, offering a common API, or even enabling development of new environments. Finally, we describe the tools used specifically for Reinforcement Learning, either by providing full algorithms, components of them, or other auxiliary functionalities.

All the listed frameworks are written for the Python programming language, although they frequently use other languages for efficient computation that the end user does not need to know. This is the de facto standard in Machine Learning and Reinforcement Learning specifically. Despite its relatively slow performance, through the use of the aforementioned libraries, all the heavy computation is off-loaded to a more efficient language that the end used does not need to use directly. 

\subsection{Neural Networks} \label{sec:frameworks-nns}

\begin{figure}
    \centering
    \includegraphics[width=\linewidth]{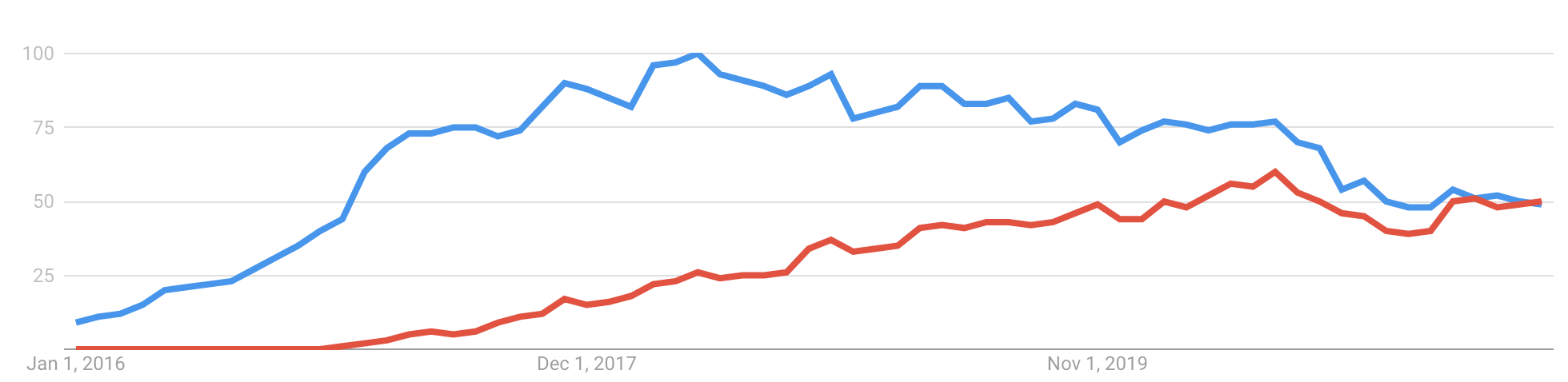}
    \caption{The relative popularity of PyTorch ({\color{red} red}) and TensorFlow ({\color{blue}blue}) in terms of search volume on Google according to Google Trends, worldwide, between 01/01/2016 and 27/07/2021.}
    \label{fig:framework-trends}
\end{figure}

While many open-source tensor computation libraries exist today, three in particular stand out. The first is \textbf{TensorFlow (TF)}~\cite{martin_abadi_tensorflow_2015} developed by Google -- originally released in 2015, it underwent major changes in 2019 when the version 2.0 was released with a new philosophy. For this reason, TF1 and TF2 are mutually incompatible and can be seen as two distinct frameworks. The main difference between them is the default handling of computation graphs. In TF1, we have to explicitly define a graph, and then run it within a session. This means that any intermediate values are hidden from the user, leading to a difficult debugging process. On the other hand, TF2 uses eager evaluation, building an implicit computation graph. This way, the developer can access the values of any tensors at any time, also in interactive mode \eg in a Jupyter Notebook~\cite{loizides_jupyter_2016}, without needing to modify the computation graph for this purpose. TensorFlow can run computations on CPU, GPU (via CUDA and ROCm), and TPU.

It is important to mention \textbf{Keras}~\cite{chollet_keras_2015}, originally developed as an independent library, is now integrated as part of TF2. It exposes a higher-level API that allows faster development at the cost of fine-tuned control over the computation graph -- this, however, can be regained by including lower-level TF2 code as custom layers. While it is primarily designed for Supervised Learning, it is possible to use it with Reinforcement Learning, particularly when using only some of its features in conjunction with TF2 code.

\textbf{PyTorch}~\cite{paszke_pytorch_2019}, developed by Facebook, was released in 2016 as a Python version of the existing library Torch, which used the Lua language. Its design is very similar to NumPy~\cite{harris_array_2020} and inspired TF2 in that it uses eagerly-executed tensors that can be used in dynamic computation graphs. PyTorch can run computation on CPU and GPU (via CUDA), and with some extra effort, TPU. Its simplicity of use, combined with performance that matches TensorFlow, led to its widespread usage, particularly in research context~\cite{he_state_2019}. 

A relatively new framework that is worth mentioning is \textbf{Jax}~\cite{bradbury_jax_2018}. It offers simple acceleration and parallelization of code with a simple interface that can nearly be used as a drop-in replacement for NumPy. It is heavily inspired by the functional programming paradigm, focuses on composable transformations of functions, notably including differentiating arbitrary functions. By itself, Jax contains efficient numerical operations on tensors, and does not explicitly include neural networks. However, libraries in its ecosystem fill that gap, notably Flax~\cite{heek_flax_2020} and Haiku~\cite{hennigan_haiku_2020}.

In the words of Andrej Karpathy, "I've been using PyTorch a few months now and I've never felt better. I have more energy. My skin is clearer. My eye sight has improved."~\cite{karpathy_ive_2017} This sentiment is also visible in the data. According to the 2020 Stack Overflow Developer Survey~\cite{stack_overflow_stack_2020}, while TensorFlow is more commonly used than PyTorch (11.5\% vs 4.6\%), it has a lower Loved score (65.2\% vs 70.5\%), and higher Dreaded score (34.8\% vs 29.5\%). According to Google Trends, despite its later release date, PyTorch matches the popularity of TensorFlow is terms of search volume, as we show on Figure~\ref{fig:framework-trends}. Overall, the common perception is that TensorFlow is more suited for deployment and industry applications, while PyTorch can be more effective for research and development. 

\subsection{Environments} \label{sec:frameworks-envs}


\begin{figure}
    \centering
    \includegraphics[width=\linewidth]{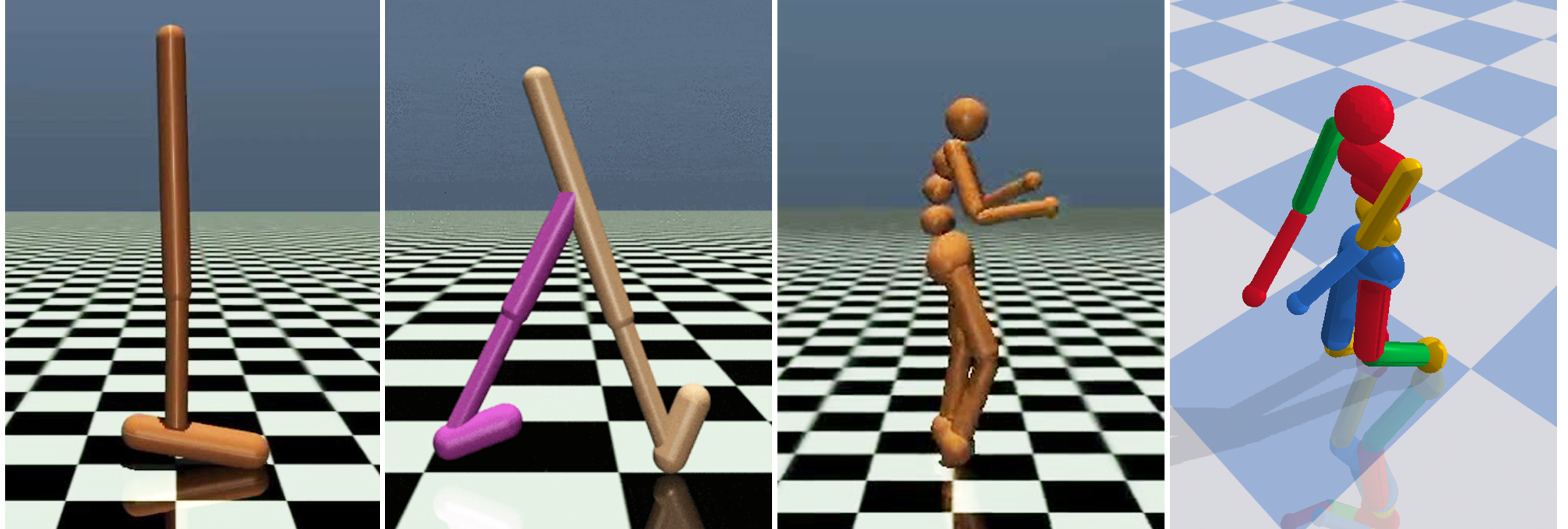}
    \caption{A visualization of different legged models of varying complexity. The agents' objective is moving each of the joints so that the overall center of mass moves forward or is balanced, while minimizing the energy expenditure. From left to right: Hopper, Walker2d, Humanoid (MuJoCo) and Humanoid (PyBullet).}
    \label{fig:legged-models}
\end{figure}

The main framework underlying nearly all modern RL research is \textbf{Gym}~\cite{brockman_openai_2016}. It contains a set of commonly used environments that serve as benchmarks for RL algorithms, together with a unified Environment API and a way to implement new environments as Python classes. This way, a single implementation of the environment can be used with multiple algorithms for easier comparison and benchmarking.

The main parts of the Gym API are the following methods:
\begin{itemize}
    \item reset() $\to$ observation
    \item step(action) $\to$ (observation, reward, done, info)
\end{itemize}
where the observation and action can be in any format agreed between the environment and the agent. The observations are usually tensors, and actions are either vectors or discrete values, but more complex, tree-like structures are also sometimes used. The reward is a single scalar value, and done is a binary flag indicating whether an episode has ended. Finally, info is a dictionary containing arbitrary additional information.
Other properties like observation\_space and action\_space exist to specify the structure of the information exchanged with the agent, but the extent to which they are required depends on the specific algorithm implementation.

Canonically, Gym only supports partially observable single-agent scenarios, corresponding to POMDPs. Supporting multiagent environments is possible to an extent without changing the abstractions. Specifically, a DecPOMDP can be represented by taking the state and action spaces to be the product spaces of all agents. The reward can then remain as a single scalar value, shared between the agents, maintaining full compatibility with the Gym API. 

The situation is more complicated when dealing with more general multiagent problems like POSGs. In this case, each agent may receive its own reward independently of others, which cannot be represented with a single scalar value. What is more, some agents might not be active throughout the episode. For that reason, frameworks like \textbf{RLlib}~\cite{liang_rllib_2018} use a modified version of a gym Environment where everything is based on Python dictionaries -- observations, actions, rewards and `done' values (\ie boolean episode termination flags) are Python dictionaries, where each agent's respective values are indexed by the name of that agent.

Other general multiagent formalisms also have their corresponding libraries. The \textbf{Petting Zoo}~\cite{terry_pettingzoo_2021} library implements the AEC formalism, and is well-suited for general multi-agent problems in which agents do not act simultaneously, but also supports simultaneous actions. Similarly to Gym, it also contains a number of standard benchmark multiagent environments. The EFG formalism is implemented in the OpenSpiel~\cite{lanctot_openspiel_2020} library, which also contains several ready board and card games.

\textbf{MuJoCo}~\cite{todorov_mujoco_2012}, which stands for Multi-Joint dynamics with Contact, is a physics simulation engine that is widely used in RL research. It can be used to design various robots whose control is then learned with RL algorithms. Certain robots are included in Gym and serve as a common benchmark for new algorithms, notably the Humanoid~\cite{tassa_synthesis_2012} which is relevant to the topic of character animation. It is worth noting that MuJoCo used to require a paid license to use, but has now become open-source. \textbf{PyBullet}~\cite{coumans_pybullet_2016} and \textbf{DART}~\cite{lee_dart_2018} are common open-source alternatives, filling the same role as a physics simulation engine, and containing many of the same environments ready to use. \textbf{NVIDIA Isaac Gym}~\cite{liang_gpu-accelerated_2018} fulfills a similar role, enabling very fast parallel simulations accelerated with GPUs.
Different legged models from MuJoCo and PyBullet can be seen in Figure~\ref{fig:legged-models}.

\textbf{ML-Agents}~\cite{juliani_unity_2020} is a plugin to the Unity game engine which exposes an API through which Python-based agents can interact with games developed in Unity. This can greatly accelerate the development of new environments, as many features from game development are available out-of-the-box, and are relevant for RL tasks. ML-Agents also contains implementations of PPO and SAC algorithms in PyTorch, with the possibility to record and train with user-made demonstrations, using BC or GAIL.

\textbf{Osim}~\cite{kidzinski_learning_2018}, based on the OpenSim framework~\cite{seth_opensim_2011}, was used in the 2017 NeurIPS "Learning to Run" challenge is a simulator with a physiologically accurate model of the human body. Its main goal is bridging the gap between biomechanics, neuroscience and computer science communities by providing a common ground for research. It contains a physics simulator, an RL environment, as well as a competition platform to compare different solutions.

Alternatively to gradient-free physical engines, further implementations such as \textbf{Nimble}~\cite{Werling2021} introduce novel ways of fast and complete differentiable rigid bodies simulations, which can be used for hard optimization problems dealing with complex contact geometries or elastic collisions. Differentiable physics~\cite{Macklin2021} and the possible combination with stochastic gradient-free methods show promising directions for research into more efficient physics engines for learning and optimization.

\subsection{Algorithm implementations} \label{sec:frameworks-algos}

In recent years, many frameworks with implementations of RL algorithms have appeared, many of them including the same algorithms, but with differences in the tricks included, or in the implementation philosophy. A common issue is that a framework is too rigid, and therefore it is difficult to customize it for arbitrary research purposes. Here we describe the most relevant frameworks, with a comparison of the algorithms they feature in Table~\ref{tab:frameworks}

One of the earliest libraries with high-quality implementations of modern RL algorithms is \textbf{OpenAI Baselines}~\cite{dhariwal_openai_2017}. Released in 2017, it contains the implementations of the most important algorithms existing at the time, including DQN, DDPG and PPO, using TensorFlow. However, at the time of writing it is in maintenance mode, which means that it is not updated with the new developments in the field. Furthermore, the implementations are considered to not be very readable and are challenging to modify or update.

For this reason, the \textbf{Stable Baselines (SB2)}~\cite{hill_stable_2018} library was developed, with re-implementations of the same algorithms with an addition of SAC and TD3, with a series of improvements to their usability, including: better documentation, tests, using custom policies, and a shared interface between all algorithms. A full comparison is included in the official GitHub repository. Stable Baselines uses TensorFlow for its algorithms, but is now in maintenance mode. The more up-to-date version is \textbf{Stable Baselines 3 (SB3)}~\cite{raffin_stable_2019} which has the same purpose as SB2, but uses PyTorch instead, and is actively updated with new features. Both SB2 and SB3 have very limited support for multiagent training.

\textbf{RLlib}~\cite{liang_rllib_2018} is built on Ray~\cite{moritz_ray_2018}, a platform for parallel computing, and because of that it can achieve high performance through efficiently parallelizing data collection. It has a large selection of both single-agent and multiagent algorithms in both PyTorch and TensorFlow (depending on the algorithm), with a wide variety of options to adjust for those algorithms. The implementations are very efficient, but the code-base is complex, and therefore difficult to understand and modify for most users.

\textbf{CleanRL}~\cite{huang_cleanrl_2020} is a library with a different design philosophy -- all implementations are contained entirely in a single file. This allows for simple customization, and can serve as a reference when re-implementing those algorithms. CleanRL uses PyTorch for its algorithm implementations. It also includes the Open RL Benchmark, which contains reproducible experiments that use the implemented algorithms on a wide range of standard RL environments.

\textbf{Dopamine}~\cite{castro_dopamine_2018} is a research framework developed by Google, with the design principles of \textit{easy experimentation}, \textit{flexible development}, \textit{compactness and reliability}, and \textit{reproducibility}. It focuses on variants of DQN, including Rainbow and several other modifications. The framework primarily uses TensorFlow, but it also contains Jax implementations of the algorithms.

\textbf{TF-Agents}~\cite{guadarrama_tf-agents_2018} is a part of the TensorFlow ecosystem dedicated to Bandits and Reinforcement Learning algorithms. It contains high-quality implementations of modern RL algorithms written in TF2. The implementations are modular and include tests, benchmarks and tutorials on how to use the library.

\textbf{Tianshou}~\cite{weng_tianshou_2021} maintained by researchers from Tsinghua University is a modular RL framework based on PyTorch. While focusing on single-agent model-free algorithms, it also supports multiagent environments, model-based algorithms and imitation learning. The implementations are flexible so that it is possible to modify them for research purposes, and very efficient due to parallelization. 

\textbf{Rlax}~\cite{budden_rlax_2020}, developed by Deepmind as part of the Jax ecosystem, takes a different approach than all the aforementioned frameworks. Instead of full algorithms, it contains building blocks that can then be used in implementing the algorithms. This includes exploration strategies, policies, update strategies and more. 

\subsection{Summary}

As we show, there are numerous libraries that can be used for creating environments and training RL algorithms, and the best choice will necessarily depend on the application. Regarding environment creation, highly specialized problems might require custom simulators, but for generic character and crowd animation problems, we recommend using the Unity engine with ML-Agents to build a gym or gym-like interface. For training, if customization of the algorithm is unnecessary, we recommend using either RLlib or Stable Baselines 3 for single-agent (skeletal animation) scenarios. If some degree of customization is necessary, Stable Baselines 3 or Tianshou are worthwhile options, with RLlib typically being too rigid. Finally, if it is necessary to introduce major changes to the typical reinforcement learning loop, it might be necessary to use a custom-written algorithm - components from Tianshou, CleanRL and Rlax can then prove to be helpful. Naturally, all of this is conditioned on the availability of the desired algorithm in a specific framework (see Table~\ref{tab:frameworks}).

%% file: tables/frameworks.tex
\begin{table*}[ht] \centering
\caption{A comparison of algorithm support between various frameworks. Legend: $\checkmark$ -- algorithm supported by the framework, $\times$ -- algorithm not supported by the framework. Multiagent refers to the capability of training in multiagent environments, with or without parameter sharing. Note that this is not a complete list of algorithms implemented by each framework, as some of them include many other, less relevant algorithms.} \label{tab:frameworks}
\begin{tabular}{cccccccccc}
Algorithm &
  \begin{tabular}[c]{@{}c@{}}OpenAI\\ Baselines\end{tabular} &
  \begin{tabular}[c]{@{}c@{}}Stable\\ Baselines\end{tabular} &
  \begin{tabular}[c]{@{}c@{}}Stable \\ Baselines 3\end{tabular} &
  RLLib &
  CleanRL &
  Dopamine &
  TF-Agents &
  Tianshou &
  ML-Agents \\ \hline
DQN        & $\checkmark$ & $\checkmark$ & $\checkmark$ & $\checkmark$ & $\checkmark$ & $\checkmark$ & $\checkmark$ & $\checkmark$ & $\times$     \\
Rainbow    & $\times$     & $\times$     & $\times$     & $\checkmark$ & $\times$     & $\checkmark$ & $\times$     & $\times$     & $\times$     \\
DDPG       & $\checkmark$ & $\checkmark$ & $\checkmark$ & $\checkmark$ & $\checkmark$ & $\times$     & $\checkmark$ & $\checkmark$ & $\times$     \\
TD3        & $\times$     & $\checkmark$ & $\checkmark$ & $\checkmark$ & $\checkmark$ & $\times$     & $\checkmark$ & $\checkmark$ & $\times$     \\
SAC        & $\times$     & $\checkmark$ & $\checkmark$ & $\checkmark$ & $\checkmark$ & $\times$     & $\checkmark$ & $\checkmark$ & $\checkmark$ \\
TRPO       & $\checkmark$ & $\checkmark$ & $\times$     & $\times$     & $\times$     & $\times$     & $\times$     & $\checkmark$ & $\times$     \\
PPO        & $\checkmark$ & $\checkmark$ & $\checkmark$ & $\checkmark$ & $\checkmark$ & $\times$     & $\checkmark$ & $\checkmark$ & $\checkmark$ \\
QMIX       & $\times$     & $\times$     & $\times$     & $\checkmark$ & $\times$     & $\times$     & $\times$     & $\times$     & $\times$     \\
BC/GAIL    & $\checkmark$ & $\checkmark$ & $\times$     & $\checkmark$ & $\times$     & $\times$     & $\times$     & $\checkmark$ & $\checkmark$ \\
Multiagent & $\times$     & $\times$     & $\times$     & $\checkmark$ & $\times$     & $\times$     & $\times$     & $\checkmark$ & $\checkmark$
\end{tabular}
\end{table*}

%% file: text/10-conclusions.tex
\section{Conclusions}

Reinforcement Learning is a rapidly growing field of Artificial Intelligence and Machine Learning, concerned with authoring intelligent behaviors by specifying their tasks, instead of describing the specific behaviors. This method is of high utility for applications in Character Animation, both for individual characters, as well as entire crowds. 

In fact, many works already use RL algorithms to create more believable or higher-quality animations. In many cases, with an appropriate simulator, it is sufficient to specify the desired task in terms of a reward function (\eg agents in a crowd heading towards a certain goal, while avoiding collisions with one another), and then train one of the state-of-the-art algorithms to obtain interesting behaviors. However, Deep Reinforcement Learning is still a relatively young field, and thus its use is not common in the industry. This is likely to change in the upcoming years.

As for the implementation, there are many resources available to significantly accelerate the development of new applications of RL. While different frameworks excel in different aspects, there exist options for diverse use cases and degrees of complexity, so that in-depth expertise in the inner workings of RL algorithms is not absolutely necessary to be able to apply them.

We anticipate significant progress on the intersection of Character Animation and Reinforcement Learning in the upcoming years. The algorithms become more and more efficient, seeing success after success in classic challenges like the games of Go and Starcraft. This, combined with the widespread usage of GPUs, will make it possible to seamlessly integrate them into typical Computer Graphics workflows.

\section*{Acknowledgement}

This work has received funding from the European Union’s Horizon 2020 research and innovation programme under the Marie Skłodowska-Curie grant agreement No 860768 (CLIPE project).